\documentclass[a4paper,11pt]{article}
\pdfoutput=1 

\usepackage{jcappub} 

\usepackage[T1]{fontenc} 
\usepackage{bm}
\usepackage{subcaption}
\usepackage[export]{adjustbox}

\usepackage{scalerel,tikz}
\usetikzlibrary{svg.path}
\definecolor{orcidlogocol}{HTML}{A6CE39}
\tikzset{orcidlogo/.pic={
 \fill[orcidlogocol] svg{M256,128c0,70.7-57.3,128-128,128C57.3,256,0,198.7,0,128C0,57.3,57.3,0,128,0C198.7,0,256,57.3,256,128z};
 \fill[white] svg{M86.3,186.2H70.9V79.1h15.4v48.4V186.2z}
 svg{M108.9,79.1h41.6c39.6,0,57,28.3,57,53.6c0,27.5-21.5,53.6-56.8,53.6h-41.8V79.1z M124.3,172.4h24.5c34.9,0,42.9-26.5,42.9-39.7c0-21.5-13.7-39.7-43.7-39.7h-23.7V172.4z}
 svg{M88.7,56.8c0,5.5-4.5,10.1-10.1,10.1c-5.6,0-10.1-4.6-10.1-10.1c0-5.6,4.5-10.1,10.1-10.1C84.2,46.7,88.7,51.3,88.7,56.8z};
}}
\newcommand\orcidicon[1]{\href{https://orcid.org/#1}{\mbox{\scalerel*{
\begin{tikzpicture}[yscale=-1,transform shape]
\pic{orcidlogo};
\end{tikzpicture}
}{|}}}}

\newcommand{\nhat}{\hat{\bm{n}}}
\newcommand{\khat}{\hat{\bm{k}}}
\newcommand{\deltaa}{\delta^{a}}
\newcommand{\deltab}{\delta^{b}}
\newcommand{\deltagtwoD}{\delta^{(2D)}}

\newcommand{\deltaatwoD}{\delta_{a}^{(2D)}}

\newcommand{\deltabtwoD}{\delta_{b}^{(2D)}}

\newcommand{\phibar}{\bar{\phi}}
\newcommand{\gbar}{\bar{g}}
\newcommand{\phibara}{\bar{\phi}^a}
\newcommand{\phibarb}{\bar{\phi}^b}
\newcommand{\chishift}{\chi_{\mathrm{shift}}}

\newcommand{\dphi}{\Delta \phi}
\newcommand{\dg}{\Delta g}
\newcommand{\dphia}{\Delta \phi^a}
\newcommand{\dphib}{\Delta \phi^b}
\newcommand{\dndz}{\frac{d n_{g}}{dz}}
\newcommand{\dnbardz}{\frac{d \bar{n}_{g}}{dz}}

\newcommand{\dNdz}{\frac{d N_{g}}{dz}}

\newcommand{\dkthreeoverpithree}{\frac{\mathrm{d}^{3} \bm{k}}{(2\pi)^3}\,}
\newcommand{\dkonethreeoverpithree}{\frac{\mathrm{d}^{3} \bm{k}_1}{(2\pi)^3}\,}
\newcommand{\dktwothreeoverpithree}{\frac{\mathrm{d}^{3} \bm{k}_2}{(2\pi)^3}\,}

\newcommand{\dxthree}{d^{3} \bm{x}\,}

\newcommand{\dchi}{d \chi \,}
\newcommand{\dchione}{d \chi_1 \,}
\newcommand{\dchitwo}{d \chi_2 \,}
\newcommand{\dk}{d k \,}
\newcommand{\dkone}{d k_1 \,}
\newcommand{\dktwo}{d k_2 \,}

\newcommand{\dnhat}{d \hat{\bm{n}} \,}
\newcommand{\kvec}{\bm{k}}
\newcommand{\xvec}{\bm{x}}

\graphicspath{{plots/}}

\title{The Impact of Anisotropic Redshift Distributions on Angular Clustering}

\author[]{Antón Baleato Lizancos~\orcidicon{0000-0002-0232-6480}}
\author[]{and Martin White~\orcidicon{0000-0001-9912-5070}}
\affiliation{Berkeley Center for Cosmological Physics, UC Berkeley, CA 94720, USA}
\affiliation{Department of Physics, University of California, Berkeley, CA 94720, USA}
\affiliation{Lawrence Berkeley National Laboratory, One Cyclotron Road, Berkeley, CA 94720, USA}

\emailAdd{a.baleatolizancos@berkeley.edu}
\emailAdd{mwhite@berkeley.edu}

\abstract{A leading way to constrain physical theories from cosmological observations is to test their predictions for the angular clustering statistics of matter tracers, a technique that is set to become ever more central with the next generation of large imaging surveys.  Interpretation of this clustering requires knowledge of the projection kernel, or the redshift distribution of the sources, and the typical assumption is an isotropic redshift distribution for the objects. However, variations in the kernel are expected across the survey footprint due to photometric variations and residual observational systematic effects.  We develop the formalism for anisotropic projection and present several limiting cases that elucidate the key aspects.  We quantify the impact of anisotropies in the redshift distribution on a general class of angular two-point statistics.  In particular, we identify a mode-coupling effect that can add power to auto-correlations, including galaxy clustering and cosmic shear, and remove it from certain cross-correlations.  If the projection anisotropy is primarily at large scales, the mode-coupling depends upon its variance as a function of redshift; furthermore, it is often of similar shape to the signal. In contrast, the cross-correlation of a field whose selection function is anisotropic with another one featuring no such variations -- such as CMB lensing -- is immune to these effects.  We discuss explicitly several special cases of the general formalism including galaxy clustering, galaxy-galaxy lensing, cosmic shear and cross-correlations with CMB lensing, and publicly release a code to compute the biases.
}

\begin{document}
\maketitle
\flushbottom

\section{Introduction}
\label{sec:intro}

The measurement of angular clustering of projected fields holds a central place in cosmology, and such measurements frequently allow us to push the cosmological frontier in sky coverage, source density and redshift \cite{ref:peebles_80,ref:peebles_93,Peacock99,Dodelson03,Dodelson20,Baumann22}.  The next generation of large imaging surveys, such as Vera Rubin LSST \cite{2012arXiv1211.0310L,LSST,ref:lsst} and Euclid \cite{ref:euclid_12,Amendola18} will ensure such analyses remain central to cosmological inference for the next decade and beyond.  With some assumptions about evolution and statistical isotropy, observations along the past light cone can be used to infer the underlying 3D clustering if the radial distribution of the signal is known.  This is typically related to the redshift distribution of a set of sources, and the general assumption is that this distribution is independent of location on the sky. 

However, observational systematic effects are expected to introduce anisotropy in the selection of sources. For instance, it is well known that the observed number density of galaxies is a modulation of the true population by variations in detector sensitivity across time and location on the focal plane, changes in the observing conditions, completeness near bright stars, extinction due to Galactic dust, deblending, or the separation of galaxies from stars; this modulation can in turn bias cosmological constraints~\cite{ref:huterer_et_al_13, ref:shafer_and_huterer_15, ref:weaverdyck_and_huterer_20}. The same systematics will also compromise characterizations of the redshift distribution of the objects, presenting a challenge to the interpretation of the observed angular pattern in terms of an underlying 3D distribution. In this paper, we investigate how the analysis needs to be changed if the assumption of isotropy in the redshift distribution is relaxed, illustrating the general formalism with several examples.

The outline of the paper is as follows.  In \S\ref{sec:formalism} we introduce the general formalism for treating anisotropic projection kernels for 2D fields.  We demonstrate the impact of such anisotropic projection in a simple ``flat sky'' calculation (\S\ref{sec:flat}) and then show that the main structure of the calculation -- extra additive and mode-coupling contributions to the clustering -- carries across to the full-sky calculation (\S\ref{sec:full}).  We give several examples and special cases in \S\ref{sec:special}, showing what the general formalism implies for galaxy clustering, cross-correlations, cosmic shear and galaxy-galaxy lensing.  Then, in \S\ref{sec:examples}, we provide a quantitative exploration of the additional contributions for fluctuations characteristic of present-day surveys.  Our conclusions are presented in \S\ref{sec:conclusions} while some technical details are relegated to appendices.

\section{Formalism}
\label{sec:formalism}

We are interested in considering a field, such as the galaxy density, that has been projected along the line-of-sight direction, such that e.g.
\begin{equation}\label{eqn:generic_projection}
    1 + \delta^{(2D)}(\nhat) = \int d\chi\ \phi(\chi,\nhat)\,\left[ 1 + \delta^{(3D)}(\chi\,\nhat) \right]
\end{equation}
where $\chi$ is the comoving radial coordinate and $\nhat$ is a unit vector on the sphere.  The projection kernel, $\phi$, is inferred from an assumed or observed redshift distribution, e.g.\ for the projected galaxy density example above $\phi\propto H\,(dN/dz)$, where $H$ is the Hubble parameter and $dN/dz$ is the redshift distribution of the sources; we shall discuss other kernels later.  The standard assumption is that $\phi$ is a function only of $\chi$, with no angular dependence.  In this paper we are interested in the case where $dN/dz$ varies with position on the sky, and hence $\phi=\phi(\chi,\nhat)$.

Throughout we shall assume that $\phi(\chi,\nhat)$ is well characterized so that we can ignore the problem of redshift-distribution errors and focus instead on the impact of a spatially-varying redshift distribution.  The issue of properly inferring the redshift distribution of a population of objects, and its uncertainties, is a complex one with a large literature.  Recent reviews of the current state-of-the-art can be found in refs.~\cite{Salvato19,Newman22}.  We assume that this process has resulted in an estimate of the number of galaxies in redshift bins, including its variation across the survey, and hence of $\phi(\chi,\nhat)$.  We denote the average of $\phi$ across the sky (i.e.\ survey) as $\bar{\phi}(\chi)$, which defines the residual through
\begin{equation}\label{eqn:def_phi_full}
    \phi(\chi,\nhat) = \bar{\phi}(\chi) + \Delta\phi(\chi,\nhat)
\end{equation}
Note that by definition $\Delta\phi$ averages to zero and $\bar{\phi}$ integrates to unity\footnote{The latter condition, which holds automatically for densities, need not hold for more general fields, e.g.\ cosmic shear.  We shall relax this condition when appropriate.}.
We shall also expand the angular dependence of our fields in spherical harmonics, e.g.
\begin{equation}
    \delta^{(2D)}(\nhat) = \sum_{\ell m}\delta_{\ell m}^{(2D)}
    Y_{\ell m}(\nhat)
    \, , \quad
    \Delta\phi(\chi,\nhat) = \sum_{\ell m}\Delta\phi_{\ell m}(\chi)
    Y_{\ell m}(\nhat) \,.
\end{equation}

We are concerned with the implications of non-zero $\Delta\phi$ on the observed auto- and cross-clustering of the projected field, $\delta^{(2D)}$.

\subsection{Flat sky}
\label{sec:flat}

Before presenting our full calculation (\S\ref{sec:full}), let us consider an algebraically simpler model that nonetheless illustrates many of the features of the full model.  Specifically imagine a flat sky, with a redshift depth\footnote{This analysis might be appropriate for medium- or narrow-band selected samples of emission line galaxies over small fields, for example Ly$\alpha$ emitter surveys (LAEs; see~\cite{ref:ouchi_et_al_20} for a review).} small enough that we may directly convert angles and redshifts into Cartesian coordinates: $(\mathbf{x}_\perp,\chi)$.  We split the selection function, $\phi(\mathbf{x})$, into an average piece, $\bar{\phi}(\chi)$, plus a fluctuation $\Delta\phi(\mathbf{x}_\perp,\chi)$.  Then
\begin{equation}
    \delta_{2D}(\mathbf{x}_\perp) = \int d\chi\ \Delta\phi(\mathbf{x}_\perp,\chi) + \int d\chi\left[ \bar{\phi}(\chi)+\Delta\phi(\mathbf{x}_\perp,\chi) \right] \delta_{3D}(\mathbf{x}_\perp,\chi)
    \, .
\label{eqn:delta2D}
\end{equation}
Conservation of the mean density requires $\int\bar{\phi}\,d\chi=1$ and $\int \Delta\phi\, d^2x_\perp=0$ or equivalently $\Delta\phi(\mathbf{k}_\perp=0,k_\parallel)=0$.  In moving to Fourier space two of the terms are straightforward Fourier transforms while the $\Delta\phi\,\delta_{3D}$ term is a convolution.  Let us denote
$X(\mathbf{x})=\Delta\phi\,\delta_{3D}$.  Then
\begin{equation}
    X(\mathbf{k}) = \left[\Delta\phi\,\delta_{3D}\right](\mathbf{k}) = \int\frac{d^3k_1\,d^3k_2}{(2\pi)^6}
    (2\pi)^3\delta^{(D)}(\mathbf{k}-\mathbf{k}_1-\mathbf{k}_2)
    \Delta\phi(\mathbf{k}_1)\delta_{3D}(\mathbf{k}_2)
\end{equation}
and if the 3D power spectrum of $\delta_{3D}$ is $P(\mathbf{k})$ then
\begin{equation}
    \left\langle X(\mathbf{k}_1)X^\star(\mathbf{k}_2) \right\rangle =
    \int\frac{d^3k}{(2\pi)^3}\ \Delta\phi(\mathbf{k}_1-\mathbf{k})\Delta\phi^\star(\mathbf{k}_2-\mathbf{k})P(\mathbf{k})\,.
\end{equation}
Note that the product in configuration space has led to a convolution in Fourier space that samples a range of $\mathbf{k}$-modes around $\mathbf{k}_1$ or $\mathbf{k}_2$.

The projection over $\chi$ in equation~\ref{eqn:delta2D} implies that $\delta_{2D}(\mathbf{k}_\perp)$, is simply the Fourier transform of each contribution evaluated at $k_\parallel=0$.  The angular power spectrum thus becomes
\begin{equation}
  C_K = \int\frac{dk_\parallel}{2\pi} P(\mathbf{k}_\perp,k_\parallel)\left|\bar{\phi}(k_\parallel)\right|^2
  + C_K^{\Delta\phi} + \int\frac{d^3k}{(2\pi)^3} P(\mathbf{k})
  \left| \Delta\phi(\mathbf{K}-\mathbf{k}) \right|^2\,,
\end{equation}
where $K=|\mathbf{k_\perp}|$ and in the last integral the $\mathbf{K}=(\mathbf{k}_\perp,0)$ is interpreted as a 3-vector with zero line-of-sight component.  The first term is the usual expression resulting from a fixed projection kernel, $\bar{\phi}$.  It has the form of a power spectrum multiplied by a line-of-sight window function.  The `additive' term,
\begin{equation}
    C_K^{\Delta\phi} = \int d\chi_1\,d\chi_2\ C_K^{\Delta\phi}(\chi_1,\chi_2) = \int d\chi_1\,d\chi_2\int d^2x_\perp e^{-i\mathbf{K}\cdot\mathbf{x}_\perp}
    \Delta\phi(\mathbf{x}_\perp,\chi_1)\Delta\phi(0,\chi_2)
    \quad ,
\end{equation}
is just the angular power spectra of the projected fluctuation, $\Delta\phi$.  It comes from modulation of the mean density by the varying projection kernel.  The final term is a `mode-coupling' term that samples from $P(\mathbf{k}_\perp,k_\parallel)$ over a range of $\mathbf{k}_\perp$ around $\mathbf{K}$.  The three contributions to $C$ come from each of the contributions to $\delta_{2D}$ squared.  The cross terms vanish because of the constraint from the mean density.

This simple, flat-sky model thus leads us to expect that we will see three contributions to the measured clustering: the signal as for a uniform $dN/dz$, an additive contribution equal to the auto-correlation of the $dN/dz$ fluctuations and a mode-coupling term that couples power at an observed scale, $K$, to that of nearby scales over a range defined by the Fourier transform of $\Delta\phi$.  All three of these contributions and their behaviors carry across to the full calculation.  This simpler model also suggests that the three contributions come from the auto-spectra of the three terms in equation~(\ref{eqn:delta2D}).  This will also have an analog in the full calculation.  Finally, these results imply that if a projected field is cross-correlated with one that does not have any uncertainty in $dN/dz$ (e.g.\ CMB lensing) the corrections vanish, or if such a field is correlated with one having mean zero (e.g.\ the shear field) the additive correction vanishes.  These implications also hold in the full case, and are discussed further in what follows.

\subsection{Beyond flat sky}
\label{sec:full}

Having built intuition for the physics underpinning the calculation, let us now repeat it in the more general spherical-sky formalism, using $\chi$ to parametrize comoving distances along a line-of-sight direction specified by the unit vector $\nhat$. The general, position-dependent selection function is still given by Eq.~\eqref{eqn:def_phi_full}. Projecting the 3D density field with this selection function, as in equation~\eqref{eqn:generic_projection}, and imposing the integral constraints $\int\bar{\phi}\,d\chi=1$ and $\int \Delta\phi\, d^2\nhat=0$, we get
\begin{equation}
    \deltagtwoD(\nhat) = \int \dchi \left[ \phibar(\chi) + \dphi(\chi, \nhat) \right] \delta(\chi, \nhat) + \int \dchi  \dphi(\chi, \nhat)\,,
\end{equation}
with spherical harmonic coefficients
\begin{align}\label{eqn:delta_lm}
    \delta^{(2D)}_{\ell m} = \int \dnhat Y_{\ell m}^{*}(\nhat) \int \dchi  \left[ \phibar(\chi) \delta(\chi, \nhat)  + \dphi(\chi, \nhat)\delta(\chi, \nhat) + \dphi(\chi, \nhat) \right]\,.
\end{align}

The first term would be the only contribution if the redshift distribution were perfectly isotropic. When it is not, the other two terms give rise to a multiplicative and an additive contribution, respectively. Let us now carefully unpack the second term, leaving the other ones to follow by analogy. It gives\footnote{
    We work in the asymmetric Fourier transform convention where
    \begin{equation}
        f(\kvec) = \int \dxthree f(\xvec) e^{- i \kvec \cdot \xvec} \quad \mathrm{and} \quad f(\xvec) = \int \dkthreeoverpithree f(\kvec) e^{ i \kvec \cdot \xvec}\,.
    \end{equation}
    Further, here and in subsequent integrals, we simplify notation by letting the dependence on comoving distance appear implicitly via the redshift, $\delta(\kvec, z) \equiv \delta(\kvec, z(\chi))$. On the other hand, $\dphi$ is only defined on the past lightcone, so its Fourier transform is fully specified by a wavevector.}:
\begin{align}\label{eqn:conv_term_partially_simplifiied}  \delta^{(2D)}_{\ell m} \supset 
    & \int \dnhat Y_{\ell m}^{*}(\nhat) \int \dchi \dphi(\chi, \nhat)\delta(\chi, \nhat) \nonumber \\
    = &  \left(4\pi\right)^2 \int \dchi \int \dkonethreeoverpithree \dktwothreeoverpithree \dphi(\kvec_1) \delta\left(\kvec_2, z\right) \nonumber \\
    & \hphantom{ \left(4\pi\right)^2 \int} \times (-1)^{m} \sum_{\ell _1 m_1} \sum_{\ell _2 m_2} i^{\ell_1 +\ell_2} G^{\ell \ell_1\ell_2}_{-m m_1 m_2} j_{\ell _1}(k_1 \chi) j_{\ell _2}(k_2 \chi) Y^{*}_{\ell _1 m_1}(\khat_1) Y^{*}_{\ell _2 m_2}(\khat_2) \,,
\end{align}
where, in going to the last line, we have used Rayleigh's plane wave expansion
\begin{equation}
    e^{i \kvec \cdot \xvec}= 4\pi \sum_{\ell m} i^{\ell} j_{\ell}(k\chi) Y_{\ell m}^{*}(\khat)Y_{\ell m}(\nhat)\,,
\end{equation}
as well as the definition of the Gaunt integral
\begin{align}\label{eqn:gaunt}
    G^{\ell_1\ell_2\ell_3}_{m_1 m_2 m_3} &\equiv \int \dnhat Y_{\ell _1 m_1}(\nhat) Y_{\ell _2 m_2}(\nhat) Y_{\ell _3 m_3}(\nhat) \nonumber \\
    & = \sqrt{\frac{(2\ell_1+1)(2\ell_2+1)(2\ell_3+1)}{4\pi}}\begin{pmatrix}\ell_1&\ell_2&\ell_3 \\ m_1&m_2&m_3 \end{pmatrix} \begin{pmatrix}\ell_1&\ell_2&\ell_3 \\ 0&0&0 \end{pmatrix} \,.
\end{align}

At this point, it will be useful to extract the spherical harmonics of each radial slice.  To do this, note that an arbitrary 3D field $g$ can be expressed as
\begin{align}\label{eqn:glm_of_r}
    g_{\ell m}(\chi)\equiv & \int \dnhat Y^*_{\ell m}(\nhat) g(\chi, \nhat)\,.
\end{align}
In appendix~\ref{sec:sFB}, we link this to the spherical Fourier-Bessel basis and use that to glean insights into the structure of the perturbations. If we have two statistically-isotropic random fields, $\deltaa$ and $\deltab$, with 3D cross-spectrum
\begin{align}
    \langle \delta^{a} \left(\kvec_1, z_1\right) \delta^{b,*}\left(\kvec_2, z_2\right) \rangle = \left(2\pi\right)^3 \delta^{(3)}_{\mathrm{D}}\left(\kvec_1 - \kvec_2\right) P_{ab}\left(k; z_1,z_2\right)\,,
\end{align}
the angular cross-spectrum of two radial slices is given by
\begin{align}\label{eqn:cls_of_delta}
    \langle \delta^{a}_{\ell _1 m_1}(\chi_1) \delta^{b,*}_{\ell _2 m_2}(\chi_2) \rangle  & = \delta_{\ell _1l_2} \delta_{m_1 m_2} \frac{2}{\pi} \int \dk k^2 j_{\ell _1}(k \chi_1) j_{\ell _1}(k\chi_2) P_{ab}\left(k; z(\chi_1),z(\chi_2)\right) \nonumber \\
    & \equiv \delta_{\ell _1l_2} \delta_{m_1 m_2} C_{\ell _1}^{ab}(\chi_1,\chi_2)\,.
\end{align}
In the literature, this sometimes goes by the name of multi-frequency angular power spectrum (MAPS~\cite{Datta07}; see also refs.~\cite{Shaw14,ref:castorina_white_18a,ref:castorina_white_18b}).

The formalism is slightly different for the $\dphi$'s, since these are fixed by whatever systematic effects are driving the variations in the $dN/dz$'s and are therefore deterministic rather than stochastic. Despite commuting with ensemble averaging, we can still define a notion of their $C_\ell$'s as
\begin{align}\label{eqn:cl_dphi_of_r}
    C_{\ell}^{\dphia \dphib}(\chi_1,\chi_2) \equiv \frac{1}{2\ell+1} \sum_m \dphia_{\ell  m}(\chi_1) \Delta \phi^{b,*}_{\ell  m}(\chi_2)\,.
\end{align}

We can now continue to simplify equation~\eqref{eqn:conv_term_partially_simplifiied}. With the toolkit we have developed, we can write
\begin{align}
    \delta^{(2D)}_{\ell m} \supset  \int \dchi  (-1)^{m} \sum_{\ell _1 m_1} \sum_{\ell _2 m_2} G^{\ell \ell_1\ell_2}_{-m m_1 m_2} \dphi_{\ell _1 m_1}(\chi) \delta_{\ell _2 m_2}(\chi)\,.
\end{align}
Then, defining
\begin{align}\label{eqn:dphi_del_contraction}
    \left\{\dphi\, \delta\right\}_{\ell m}(\chi) \equiv   (-1)^{m} \sum_{\ell _1 m_1} \sum_{\ell _2 m_2} G^{\ell \ell_1\ell_2}_{-m m_1 m_2} \dphi_{\ell _1 m_1}(\chi) \delta_{\ell _2 m_2}(\chi)\,,
\end{align}
and proceeding similarly for the other terms in equation~\ref{eqn:delta_lm}, we find
\begin{align}\label{eqn:pert_delta}
    \delta^{(2D)}_{\ell m} = \int \dchi   \left[ \phibar(\chi)\delta_{\ell m}(\chi) + \dphi_{\ell m}(\chi)  + \left\{\dphi \, \delta\right\}_{\ell m}(\chi) \right]\,.
\end{align}
This integral can be regarded as a sum of contributions from spherical shells positioned at increasing distance from the observer. The first two terms are responsible for projecting the 3D anisotropy in $\delta$ and $\dphi$ onto the shells, while the last term is associated with the coupling of the angular momenta of the two fields.

Consider, in turn, the angular cross-correlation of two projected overdensity fields, $\deltaatwoD$ and $\deltabtwoD$, each with its own radial selection function and associated perturbation. On the full sky, the total angular power spectrum is
\begin{align}
    T_{\ell}^{ab} =&  \frac{1}{2\ell+1} \sum_m \langle \delta^{(2D)}_{a,\ell m} \delta^{(2D),*}_{b,\ell m} \rangle \,.
\end{align}
All the cross-terms in this expression vanish, some because the $\delta$'s have mean zero by definition, others because they entail a coupling of $\phibar$ with $\dphi$ -- since the former is isotropic, there can only be a contribution from $\dphi_{00}$, and this is zero by construction (though see appendix~\ref{appendix:linear_terms} for a generalization to the case where it is not) -- and we are left with the sum of three auto-spectra:
\begin{equation}\label{eqn:total_cls}
    T_\ell^{ab} = \int d\chi_1\,d\chi_2\ \left[ U^{ab}_\ell(\chi_1,\chi_2) + A^{ab}_\ell(\chi_1,\chi_2) + R^{ab}_\ell(\chi_1,\chi_2) \right]
\end{equation}
where the `uniform' and `additive' contributions are
\begin{equation}
    U^{ab}_\ell =  \phibara(\chi_1)\phibarb(\chi_2) C^{ab}_\ell(\chi_1,\chi_2)
    \quad \mathrm{and} \quad
    A^{ab}_\ell =  C_\ell^{\dphia \dphib}(\chi_1,\chi_2)\,,
\end{equation}
while the mode-coupling contribution is 
\begin{align}\label{eqn:Rab_full}
    R^{ab}_\ell &= (2\ell+1)^{-1}\sum_m \left\langle
    \left\{ \dphia\,\delta^{a} \right\}_{\ell m}(\chi_1)
    \{ \dphib\,\delta^{b} \}_{\ell m}^{\star}(\chi_2)
    \right\rangle \\
    &= (2\ell+1)^{-1}\sum_m
    \sum_{123} G^{\ell \ell_1 \ell_2}_{-m m_1 m_2} G^{\ell\ell_1\ell_3}_{-m m_1 m_3}
    \Delta\phi^{a}_{\ell_2 m_2}(\chi_1) \Delta\phi^{b,\star}_{\ell_3 m_3}(\chi_2)
    C^{ab}_{\ell_1}(\chi_1,\chi_2) \\
    & = \sum_{L } M^{\dphia \dphib}_{\ell L}(\chi_1,\chi_2) C_{L }^{ab}(\chi_1,\chi_2) \,.
\end{align}
In going to the last line, we used the definitions of the $C_\ell$'s in~\eqref{eqn:cls_of_delta} and~\eqref{eqn:cl_dphi_of_r}, and harnessed the analogy with the mode-coupling induced by a mask (see, e.g.\ ref.~\cite{ref:hivon_et_al_02}) by defining
\begin{align}\label{eqn:mode_coupling_matrix}
    M^{\dphia \dphib}_{\ell  L}(\chi_1,\chi_2) \equiv \frac{2L+1}{4\pi} \sum_{\ell '} (2\ell'+1) \begin{pmatrix}\ell&L&\ell' \\ 0&0&0 \end{pmatrix}^2 C_{\ell '}^{\dphia \dphib}(\chi_1,\chi_2)\,,
\end{align}
using the identity
\begin{align}
    \sum_{m m_1} G^{\ell \ell_1 \ell_2}_{-m m_1 m_2} G^{\ell \ell_1 \ell_3}_{-m m_2 m_3} 
    &= \delta_{\ell_2 \ell_3} \delta_{m_2 m_3}\frac{(2\ell+1)(2\ell_1+1)}{4\pi}
    \begin{pmatrix} \ell & \ell_1 & \ell_2 \\ 0&0&0 \end{pmatrix}^2 \,.
\end{align}
In the next section, we will use an analytically-tractable toy model to understand the structure of this mode coupling and derive a simple and accurate approximation valid in most cases of interest. In addition, in appendix~\ref{sec:evaluating_integrals}, we explain how to evaluate the integrals in equation~\eqref{eqn:total_cls} efficiently.

Our formalism reveals two important insights. First, we learn that, as long as the mean is correctly characterized, the cross-correlation of a field with uncertain $dN/dz$ with another one with no such uncertainty is completely unbiased\footnote{Contrast this with the cross-correlation of a masked field on the sphere with another one covering the full sky. In that case, the result needs to be corrected by a factor of $f_{\mathrm{sky}}$. The difference is due to the fact that coupling $\Delta \phi$ with an isotropic $\delta$ can only depend on the monopole of $\Delta \phi$, which is zero by construction. On the other hand, the sky-mean of a survey mask is precisely equal to $f_{\mathrm{sky}}$. The analogy becomes more explicit in equation~\eqref{eqn:Q_simplified}, where we allow $\dphi$ to have a non-zero monopole.}. Second, despite $\dphi$ being in general anisotropic, the biases depend only on the diagonal component of the perturbation's angular spectrum, namely $C_{\ell}^{\dphia \dphib}(\chi_1,\chi_2)$. This suggests that an estimator for the diagonal elements of $C_{\ell}^{\dphia \dphib}$ could potentially be built from measurements of the off-diagonal elements of the angular power spectrum, $T_{\ell \ell'}^{ab}$, and used to mitigate the effects we have described. We defer a more detailed exploration of this approach to future work.

\subsection{The shape of the mode-coupling integral}
\label{sec:toy_model}

In order to better understand the implications of the formalism above, let us consider a very simplified example where $\Delta\phi$ contains only a single, very large angular scale mode.  By working through this case we will be able to see how the shape of the mode-coupling contribution arises and under what conditions it mirrors that of the cosmological signal. Moreover, this simplified calculation will pave the way to a simple yet accurate analytic approximation that will be valid much more generally, in any situation where clustering is measured on smaller (angular) scales than the anisotropy of $\phi$.

Assume the shift in the mean redshift of the distribution is small compared to the width, so we can approximate
\begin{align}
    \phi(\chi, \nhat) = \frac{H(\chi)}{c}\dndz(\chi, \nhat)
    &= \frac{H(\chi)}{c} \dnbardz\left(\chi - \chi_{\mathrm{shift}}(\nhat)\right) \\
    &\approx \frac{H(\chi)}{c}\left[\dnbardz(\chi) - \chi_{\mathrm{shift}}(\nhat) \frac{d^2 \bar{n}}{d\chi dz}(\chi)\right] \,.
\end{align}
Identifying the first term with the fiducial selection function, we can isolate the perturbation
\begin{equation}
    \dphi(\chi, \nhat) =  - \chishift(\nhat) \Psi(\chi) 
\end{equation}
where we have defined $\Psi=(H/c)\, d^2 \bar{n}/(d\chi\, dz)$ having dimensions of inverse length squared.
For pedagogical purposes, consider
\begin{equation}
    \chishift(\nhat) = \epsilon \,\Re \left\{Y_{11}(\nhat)\right\}\,,
\end{equation}
where $\epsilon$ is a small constant with units of distance. It follows that
\begin{equation}
    \dphi_{\ell m}(\chi) = - \frac{1}{2}\delta_{\ell  1} \left(\delta_{m 1} - \delta_{m -1}\right) \epsilon \Psi(\chi) \,,
\end{equation}
so our approximation will be valid whenever $\epsilon H(\chi)/c\ll 1$. We can use this to calculate
\begin{equation}
    C^{\dphia \dphib}_{\ell}(\chi_1,\chi_2) = \frac{1}{6} \delta_{\ell  1} \frac{\epsilon_a \epsilon_b}{c^2} \Psi^a(\chi_1) \Psi^b(\chi_2) \,.
\end{equation}

The additive bias is therefore only present at $\ell=1$, the mode where we seeded the perturbation:
\begin{align}\label{eqn:additive_integral_toy_model}
    T_{\ell}^{ab} \supset \int d\chi_1\,d\chi_2 C_\ell^{\dphia \dphib}(\chi_1,\chi_2) = \frac{1}{6}\delta_{\ell  1} \frac{\epsilon_a \epsilon_b}{c^2}\int d\chi_1\, \Psi^a(\chi_1) \int d\chi_2\, \Psi^b(\chi_2) \,.
\end{align}

For definiteness, let us consider the specific example of a Gaussian redshift distribution with width $\sigma_0=0.06$ centered at $z_0=0.59$, and variations about the mean redshift with amplitude $\epsilon=0.3\times10^{-2}$. As we will see in greater detail in \S\ref{sec:example_gc}, our choices are inspired by DES's \textsc{RedMaGiC} sample~\cite{ref:des_y3_cosmo}. We also adopt a galaxy power spectrum appropriate for this sample and described in that same section. Translated to distance units (in the Planck 2018~\cite{ref:planck_18_legacy,ref:planck_params_18} cosmology), these parameter values correspond to $\chi_{0}=2241$\,Mpc, $\sigma_0=187$\,Mpc and $\epsilon=9.5$\,Mpc.

\begin{figure}
    \centering
    \includegraphics[width=0.7\textwidth]{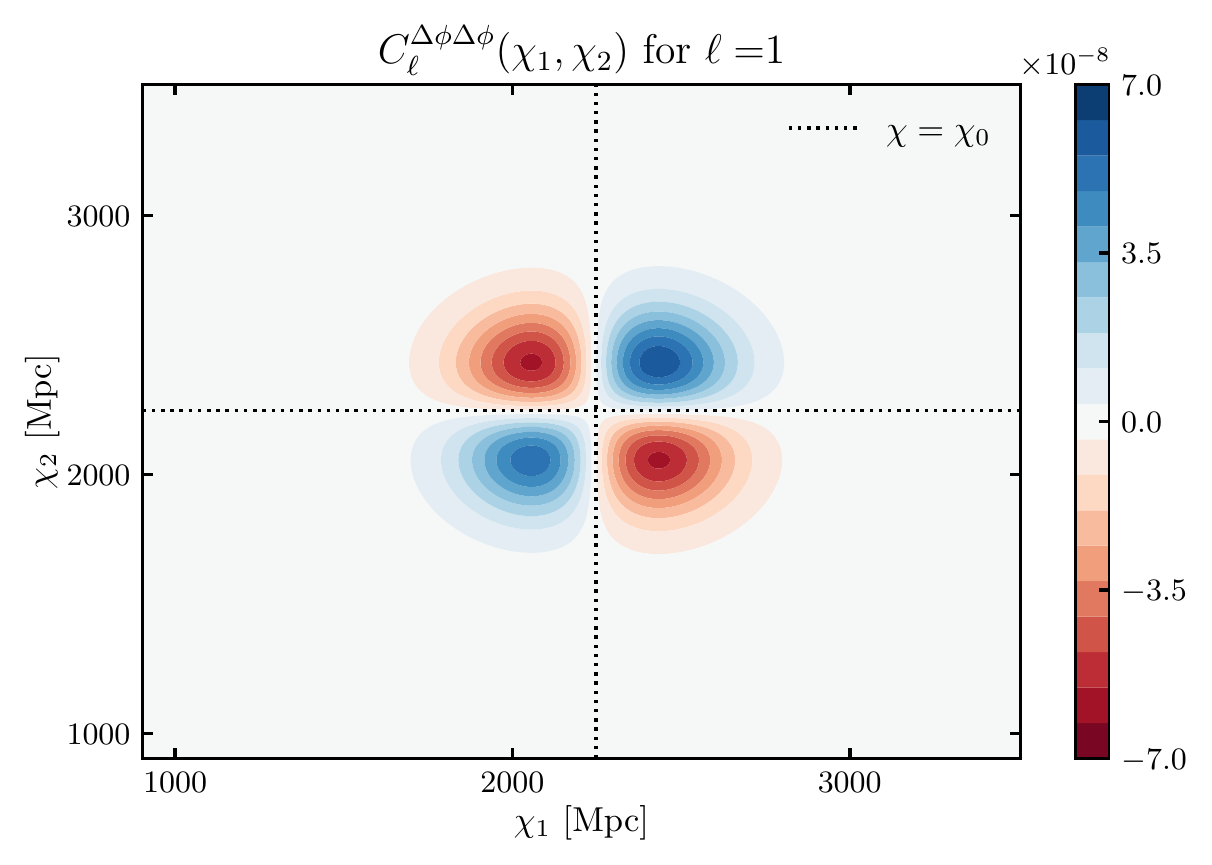}
    \caption{Multi-frequency angular power spectrum (MAPS) for our idealized example, where perturbations in the mean redshift of the distribution are seeded at $\ell=1$. In this example, $\ell=1$ is the only scale containing any structure at all. Dotted lines mark the comoving distance to the center of the galaxy distribution, $\chi_{0}$. Because the sign of $\dphi$ changes either side of $\chi_{0}$, so does the sign of the MAPS. And since the additive bias comes from an integral of the MAPS over all of $\chi_1$ and $\chi_2$, cancellations make this contribution negligibly small.}
    \label{fig:ClDphiChi1Chi2_toymodel}
\end{figure}

The integrand of equation~\eqref{eqn:additive_integral_toy_model} can be visualized in figure~\ref{fig:ClDphiChi1Chi2_toymodel}. Note the extensive cancellations between regions where the integrand takes on opposite signs, leading us to expect that the additive bias will ultimately be very small. This can be seen more explicitly if we integrate by parts:
\begin{align}
    \int_{\chi=0}^{\infty} d\chi\, \left\{H\frac{d^2 \bar{n}}{d\chi dz}\right\}(\chi) = \left[H \frac{d \bar{n}}{dz} \right]_{\chi=0}^{\infty} - \int_{\chi=0}^{\infty} d\chi\, \left\{ \frac{dH}{d\chi}\frac{d \bar{n}}{ dz}\right\}(\chi)  \,.
\end{align}
The first term on the right vanishes because the $dN/dz$ is zero at the boundary, and the second term is very small because $H$ varies slowly over a typical redshift distribution -- in the limit that $H$ is constant over this range, the integrals are exactly zero. The narrower the mean $dN/dz$, the smaller this contribution will be. For the scenario at hand, the additive bias is consistent with zero up to numerical error.

Meanwhile, the mode-coupling matrix also takes a transparent form 
\begin{align}
    M_{\ell L }^{\dphia \dphib}(\chi_1,\chi_2)=  (2L+1)  \begin{pmatrix}\ell\,&L&\,1 \\ 0&0&0 \end{pmatrix}^2 \frac{\epsilon_a \epsilon_b}{8\pi c^2}  \Psi^a(\chi_1)\Psi^b(\chi_2) \,,
\end{align}
with the triangle conditions of the $3j$ symbol imposing the simplest of off-diagonal couplings,
\begin{align}
    \begin{pmatrix}\ell\,&L&\,1 \\ 0&0&0 \end{pmatrix}^2 &= \delta_{L\,(\ell-1)} \begin{pmatrix}\ell\,&\ell-1&\,1 \\ 0&0&0 \end{pmatrix}^2  + \delta_{L\,(\ell+1)}\begin{pmatrix}\ell\,&\ell+1&\,1 \\ 0&0&0 \end{pmatrix}^2  \\
    &= \delta_{L\,(\ell-1)}\frac{\ell}{(2\ell+1)(2\ell-1)} + \delta_{L\,(\ell+1)} \frac{\ell+1}{(2\ell+3)(2\ell+1)} \, .
\end{align}
The range of  $\ell$s over which $C^{\dphia \dphib}_{\ell}(\chi_1,\chi_2)$ has support determines the width of the convolution kernel to which $P_{ab}$ is subjected. Had $\dphi$ had structure on smaller angular scales, the mode-coupling would have involved a wider range of $L$'s.  With the mode-coupling matrix above, the integrand of the mode-coupling bias becomes
\begin{align}
    R^{ab}_{\ell}
    &= \sum_{L} M_{\ell L }^{\dphia \dphib}(\chi_1,\chi_2) C_{L}^{ab}(\chi_1,\chi_2) \nonumber \\
    &= \frac{\epsilon_a \epsilon_b}{8\pi c^2} \Psi^a\Psi^b \left[(2\ell-1)C^{ab}_{\ell-1} \begin{pmatrix}\ell\,&\ell-1&\,1 \\ 0&0&0 \end{pmatrix}^2 + (2\ell+3)C^{ab}_{\ell+1} \begin{pmatrix}\ell\,&\ell+1&\,1 \\ 0&0&0 \end{pmatrix}^2\right]\,,
\end{align}
giving
\begin{align}
    T_{\ell}^{ab} &\supset  \int d\chi_1\,d\chi_2\ R^{ab}_\ell(\chi_1,\chi_2)
    \nonumber \\
    &=  \frac{\epsilon_a \epsilon_b}{8\pi c^2} (2\ell-1) \begin{pmatrix}\ell\,&\ell-1&\,1 \\ 0&0&0 \end{pmatrix}^2 \int d\chi_1\, \Psi^a(\chi_1) \int d\chi_2\, \Psi^b(\chi_2)\ C^{ab}_{\ell-1}(\chi_1,\chi_2) \nonumber \\
    &+ \frac{\epsilon_a \epsilon_b}{8\pi c^2} (2\ell+3) \begin{pmatrix}\ell\,&\ell+1&\,1 \\ 0&0&0 \end{pmatrix}^2 \int d\chi_1\, \Psi^a(\chi_1) \int d\chi_2\, \Psi^b(\chi_2)\ C^{ab}_{\ell+1}(\chi_1,\chi_2)\nonumber \\
    &\approx  \frac{\epsilon_a \epsilon_b}{8\pi c^2} \int \frac{d\chi}{\chi^{2}} \Psi^a(\chi)\Psi^b(\chi)  \nonumber \\
    & \hphantom{\epsilon_a \frac{(2\ell+1)}{2\pi^2}} \times\Bigg\{ (2\ell-1) \begin{pmatrix}\ell\,&\ell-1&\,1 \\ 0&0&0 \end{pmatrix}^2 P_{ab}\left(\frac{\ell-1/2}{\chi}; z\right)  \nonumber \\
    & \hphantom{\epsilon_a \frac{(2\ell+1)}{2\pi^2} \times\Bigg\{}+  (2\ell+3) \begin{pmatrix}\ell\,&\ell+1&\,1 \\ 0&0&0 \end{pmatrix}^2 P_{ab}\left(\frac{\ell+3/2}{\chi}; z\right) \Bigg\}
    \,.
\end{align}
In the second equality, we have used equation \eqref{eqn:limber_Clab}, the Limber approximation for $C_\ell^{ab}$. When we evaluate this expression, we find it to be in very good agreement with a full calculation of equation~\eqref{eqn:Rab_full} using simulated, perturbed distributions, underestimating it by only $1\%$ below $\ell\lesssim 2000$.

The advantage of this analytic route, however, is that it sheds light on how the mode-coupling bias depends on the characteristics of the problem. Except on the very largest angular scales we can approximate $P_{ab}\left(\frac{\ell+3/2}{\chi}; z\right)\approx P_{ab}\left(\frac{\ell-1/2}{\chi}; z\right)\approx P_{ab}\left(\frac{\ell+1/2}{\chi}; z\right)$ with an error of $\mathcal{O}(\ell^{-1})$, so
\begin{align}
    \Delta T_{\ell\gg1}^{ab} \approx  &  \frac{\epsilon_a \epsilon_b}{8\pi c^2} \left\{ (2\ell-1) \begin{pmatrix}\ell\,&\ell-1&\,1 \\ 0&0&0 \end{pmatrix}^2 +  (2\ell+3) \begin{pmatrix}\ell\,&\ell+1&\,1 \\ 0&0&0 \end{pmatrix}^2\right\}\nonumber\\
    & \times \int \frac{d\chi}{\chi^2}\ \Psi^a(\chi)\Psi^b(\chi)\  P_{ab}\left(\frac{\ell+1/2}{\chi}; z\right)  \,.
\end{align}
Then, using the identity~\cite{ref:varshalovich_book},
\begin{align}\label{eqn:3j_sum_identity}
    \sum^{\ell_1+\ell_2}_{\ell_3=|\ell_1-\ell_2|} (2\ell_3+1) \begin{pmatrix}\ell_1\,&\ell_2&\,\ell_3 \\ 0&0&0 \end{pmatrix}^2 = 1\,,
\end{align}
the term in $\{\cdots\}$ above becomes 1 and we obtain
\begin{align}\label{eqn:toy_model_convbias_result}
    \Delta T_{\ell\gg1}^{ab} \approx &  \frac{\epsilon_a \epsilon_b}{8\pi c^2}  \int \frac{d\chi}{\chi^2}\ \Psi^a(\chi)\Psi^b(\chi)\  P_{ab}\left(\frac{\ell+1/2}{\chi}; z\right) \nonumber \\
    \equiv & \int d\chi\,\left[\frac{\phibara(\chi)\phibarb(\chi)}{\chi^2}\right] f_a(\chi)f_b(\chi)  P_{ab}\left(\frac{\ell+1/2}{\chi}; z\right) \,.
\end{align}
The functions we have defined in the last line are there to facilitate comparison with the cosmological signal, equation~\eqref{eqn:signal_cls}.  Compared to the signal, the bias integrand is suppressed by two factors of
\begin{align}\label{eqn:f_toy_model}
    f_x(\chi) \equiv (8\pi)^{-1/2} (\chi-\chi^x_0)  \frac{\epsilon_x}{(\sigma^x_0)^2}\,.
\end{align}
This tells us that the mode-coupling bias to the power spectrum scales as $\Delta T_{\ell} \propto \epsilon^2/\sigma_0^4$, where $\epsilon$ parametrizes the amplitude of the $\phi(\chi)$ variations, and $\sigma_0$ is the width of the mean distribution. Samples with broad redshift distributions are therefore more robust against bias (though remember the signal amplitude also depends upon $\sigma_0$).

\begin{figure}
    \centering 
    \includegraphics[scale=0.75]{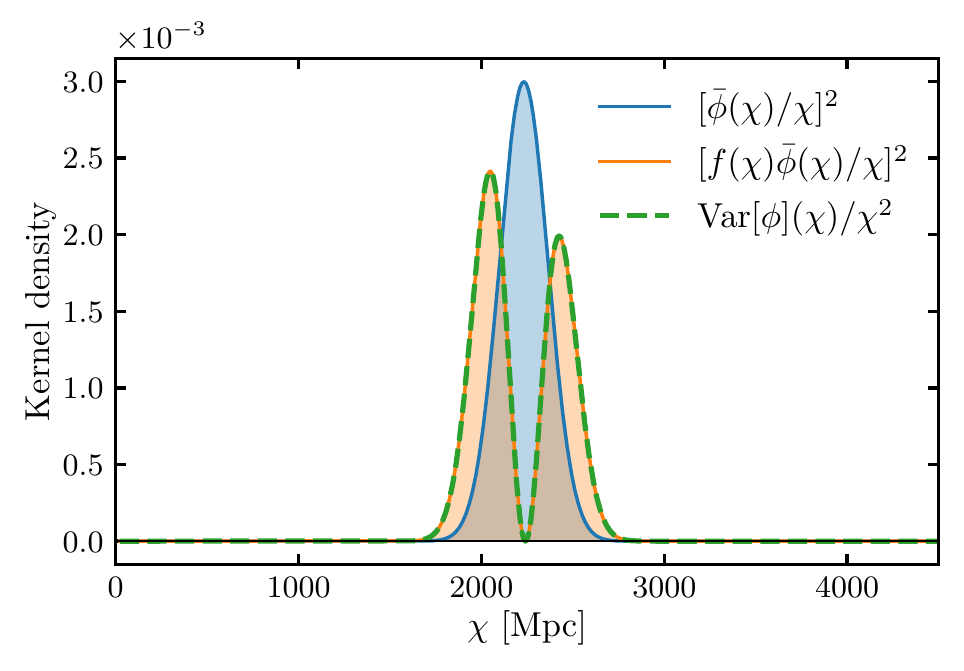}
    \includegraphics[scale=0.75]{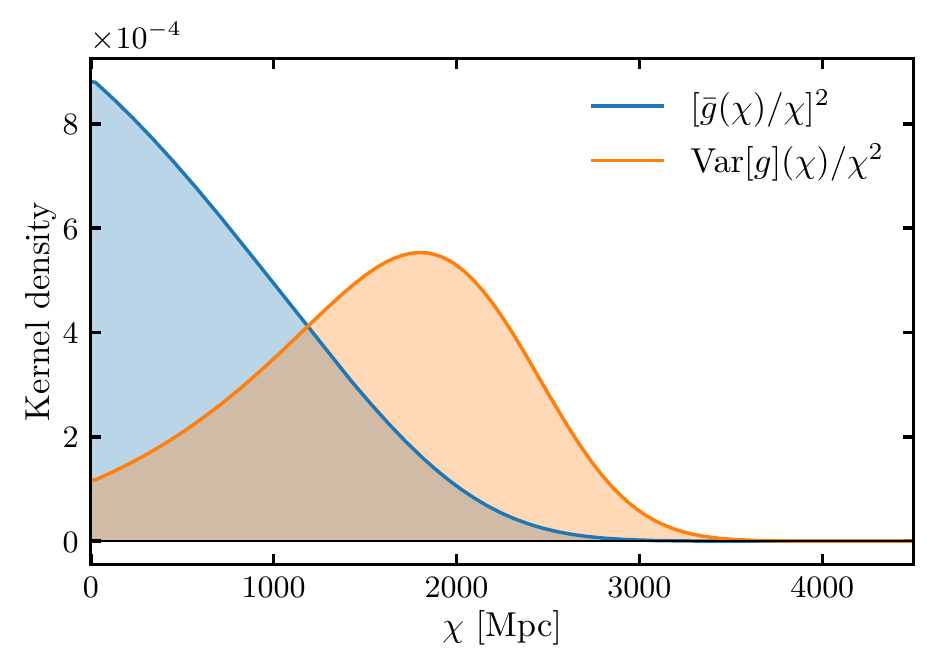} 
    \caption{Kernel densities involved in the standard angular power spectrum (blue) and the mode-coupling term in the $\ell\gg 1$ limit. The left panel pertains to our toy example of galaxy clustering, so we show the bias kernel in two different approximations (orange and green, dashed) which are practically indistinguishable from each other. Integrating the blue or orange kernels against $P^{gg}(k\sim\ell/\chi;z)$ yields very similar results, because they are probing similar effective scales and redshifts. Consequently, at large $\ell$, the bias to galaxy clustering auto-spectra has very similar shape to the signal. On the other hand, in the case of cosmic shear (the right panel being a typical such example) the effective scales and redshifts probed by the signal and bias kernels differ significantly, so we expect the shape of the two integrals to deviate accordingly.}
    \label{fig:toy_model_kernels}
\end{figure}

Moreover, the integration kernel in equation \eqref{eqn:toy_model_convbias_result} is independent of $\ell$ and probes very similar effective scales and redshifts of the 3D power spectrum, $P_{ab}(k\sim \ell/\chi;z)$, as the standard kernel in equation \eqref{eqn:signal_cls} -- see the left panel of figure~\ref{fig:toy_model_kernels}. We thus expect that, in this regime where $\ell\gg1$, the mode-coupling bias will have the same shape as the unbiased angular power spectrum, only differing from it by an $\ell$-independent amplitude factor.

This qualitative behaviour should also hold in more realistic scenarios as long as $C_{\ell'}^{\dphi}$ only has support at relatively low $\ell'$, and we are looking at much smaller scales in the angular power spectrum. When this is the case, $\ell\gg\ell'$, and the triangle condition imposes $\ell\approx L\gg\ell'$. Taking as our starting point the Limber-approximated expression in equation~\eqref{eqn:modecoupling_bias_in_limber}, we can write
\begin{align}
    \Delta T_{\ell\gg1}^{ab} \approx & \int \frac{\dchi}{\chi^2} \sum_{L } M^{\dphia \dphib}_{\ell  L} (\chi,\chi) P_{ab}\left(\frac{L+1/2}{\chi}; z\right) \\
     \approx & \int \frac{\dchi}{\chi^2} P_{ab}\left(\frac{\ell+1/2}{\chi}; z\right) \sum_{L } M^{\dphia \dphib}_{\ell  L} (\chi,\chi)\nonumber\\
      \approx &  \int \frac{\dchi}{\chi^2} P_{ab}\left(\frac{\ell+1/2}{\chi}; z\right) \sum_{\ell '} \frac{(2\ell'+1)}{4\pi} C_{\ell '}^{\dphia \dphib}(\chi) \sum_{L } (2L+1) \begin{pmatrix}\ell&L&\ell' \\ 0&0&0 \end{pmatrix}^2\,.
\end{align}
In going to the last line, we just used the definition of the mode-coupling matrix, equation~\eqref{eqn:mode_coupling_matrix}. To simplify this expression further, notice that the covariance of the perturbations across a comoving distance slice is
\begin{align}
    \mathrm{Cov}[\phi^{a}, \phi^{b}](\chi) = \mathrm{Cov}[\dphi^{a}, \dphi^{b}](\chi)  = \sum_{\ell}\frac{(2\ell+1)}{4\pi} C_{\ell}^{\dphia \dphib}(\chi) \,.
\end{align}
Finally, using identity~\eqref{eqn:3j_sum_identity} to do the sum over $L$, we arrive at the very simple expression
\begin{align}\label{eqn:simplified_mode_coupling_bias}
    \Delta T_{\ell\gg1}^{ab} \approx  \int \dchi \frac{\mathrm{Cov}[\phi^{a}, \phi^{b}](\chi)}{\chi^2} P_{ab}\left(\frac{\ell+1/2}{\chi}; z\right) \,.
\end{align}

In hindsight, this motivates equation~\eqref{eqn:f_toy_model}: $[f(\chi)\phibar(\chi)]^2$ is a remarkably good approximation to $\mathrm{Var}[\phi](\chi)$ -- see the left panel of figure~\ref{fig:toy_model_kernels} -- which is why it produced such a good approximation to the full result in the context of our toy model\footnote{Equation~\eqref{eqn:simplified_mode_coupling_bias} leads to slightly better agreement with simulations -- residuals are below $0.1\%$ in this case on scales $\ell\leq 2000$ -- because measuring the variance directly on each $\chi$-slice can capture the effect of simulation artefacts like leakage of power between angular scales.}. By the same token as above, we expect the shape of the mode-coupling contribution to the galaxy clustering power spectrum to track the signal, though the statement is now more general. In \S\ref{sec:example_gc}, figure~\ref{fig:analytic_approx}, we verify this insight and show that the expression is indeed an excellent approximation to the full calculation on scales smaller than those where there is significant projection anisotropy. The expression above also suggests that the shape and amplitude of this term relative to the galaxy clustering signal is independent of the underlying power spectrum of the objects and their distance to the observer.

Equation~\eqref{eqn:simplified_mode_coupling_bias} is in fact very general. It applies whenever the separation of angular scales is respected, even in cases where the shape of $\mathrm{Cov}[\Phi^{a}, \Phi^{b}](\chi)$ is very different from that of $\bar{\Phi}^{a}(\chi)\bar{\Phi}^{b}(\chi)$ (where $\Phi$ is now some general selection function) as is the case for example with cosmic shear or galaxy-galaxy lensing. In these cases, the various $\ell/\chi$ are weighted differently in the two integrals, and ultimately the shape of the mode-coupling term deviates from that of the signal. The difference in kernels is illustrated in the right panel of figure~\ref{fig:toy_model_kernels} for cosmic shear, and discussed further in the coming sections.

\section{Special cases}
\label{sec:special}

The formalism above is general enough that it encompasses both auto- and cross-correlations between two projected fields.  To elucidate the implications, in the following sections we tackle some special cases of particular observational interest. The full results simplify in these cases.

\subsection{Cross-correlations}

By far the easiest scenario is when at least one of the fields has a precisely isotropic redshift distribution or weight, such as occurs for example with CMB lensing. As we noted above, the three terms in equation~\eqref{eqn:total_cls} come from the auto-correlations of the $\Delta\phi$, $\bar{\phi}\,\delta$ and $\Delta\phi\,\delta$ contributions to $\delta^{(2D)}$.\footnote{This assumes that the fiducial $dN/dz$ is set to the footprint mean, as is most often done. When this is not the case, a multiplicative bias is possible, as explained in appendix~\ref{appendix:linear_terms}.}  This means that in cross correlation, if one of the fields has no $\Delta\phi$ then only the `cosmological' signal remains.  Both the additive and mode-coupling corrections vanish.

On the other hand, when the two fields being cross-correlated have projection anisotropy, a mode-coupling contribution is possible. A particularly likely and interesting scenario is when this projection anisotropy is correlated across the two fields. In \S\ref{sec:kernel_implications}, we explain that this contribution will be negative (or positive if the anisotropy is anti-correlated across fields). Then, in $\S$\ref{sec:example_gc}, we quantify its expected impact on current surveys.

\subsection{Cosmic shear}

A second case, where at least one of the fields has mean zero, retains the cosmological signal and the mode-coupling term.  However, in this case the additive contribution, $C_\ell^{\Delta\phi}$, vanishes.  Let us now show this explicitly for the case of cosmic shear.

Absent variations in the redshift distribution of the source galaxies the lensing convergence is given by\footnote{We assume a spatially-flat Universe and work in the Born approximation throughout.}
\begin{equation}
    \kappa(\nhat) =  \int d\chi\ \gbar(\chi)\, \delta(\chi, \nhat)\,,
\end{equation}
where the lens efficiency kernel is defined as
\begin{equation}\label{eqn:gbar_def}
    \gbar(\chi) \equiv \frac{3}{2} \Omega_{m,0} \frac{H_0^2}{c^2} \frac{\chi}{a(\chi)}\int_{\chi}^{\infty} d\chi_s \frac{H(\chi_s)}{c}  \frac{(\chi_s - \chi)}{\chi_s} \dnbardz(\chi_s)  \,.
\end{equation}
As with galaxy clustering, the redshift distribution of source galaxies is typically normalized such that $\int dz (d\bar{n}_g/dz)=1$. If required, convergence can then be related to shear using the Kaiser-Squires method~\cite{ref:kaiser_squires_93}.

Suppose now that the photometry varies across the sky, so that we have fluctuations around the mean redshift distribution of the source galaxy sample in different sky locations.  We can introduce $\nhat$ dependence in $d\bar{n}_g/dz$ and hence $g$ in equation~\eqref{eqn:gbar_def}, such that
\begin{equation}
    g(\chi, \nhat)\equiv \gbar(\chi) + \Delta g(\chi, \nhat)\,.
\end{equation}
As in previous sections, we assume that $\gbar$ accurately captures the sky mean, so the monopole of $\Delta g$ vanishes. In presence of these perturbations, the convergence becomes
\begin{equation}
    \kappa(\nhat) = \int d\chi \left[\gbar(\chi) + \Delta g(\chi, \nhat) \right] \delta(\chi, \nhat)\,.
\end{equation}
Notice that the perturbed convergence has no additive contribution from just $\Delta g$. In the case of galaxy clustering, this contribution appeared because any non-cosmological variation in the number of sampled galaxies across the footprint can be mistaken for the signal of interest. By contrast, in the context of cosmic shear, a variation in the number of source galaxies only affects the number of measurements available to extract the shear signal (thus imprinting inhomogeneity in the shape noise across the sky) but does not add spurious lensing signal. 

Proceeding by analogy with \S\ref{sec:full}, we have
\begin{align}
    \kappa_{\ell m} = \int \dchi   \left[ \gbar(\chi)\delta_{\ell m}(\chi) + \left\{\dg \delta\right\}_{\ell m}(\chi) \right]\,,
\end{align}
with the difference that $\delta$ now denotes the matter instead of the galaxy overdensity. Indeed, the only extra effect is a coupling of the cosmological anisotropy of $\delta$ with the newly-induced anisotropy in the lens efficiency kernel.

The impact on the cross-correlation between various tomographic shear bins can be obtained by exact analogy with \S\ref{sec:full}, simply replacing $\phi\rightarrow g$. On the full sky, the total angular cross-spectrum between bins $i$ and $j$ is
\begin{align}
    T_{\ell}^{ij} =&  \frac{1}{2\ell+1} \sum_m \langle \kappa^{i}_{\ell m}\kappa^{j,*}_{\ell m} \rangle \nonumber\\
    = & \int d\chi_1\,d\chi_2 \left[ \bar{g}^i(\chi_1)\bar{g}^j(\chi_2) C^{ij}_\ell(\chi_1,\chi_2) + \sum_{L } M^{\dg^i \dg^j}_{\ell L}(\chi_1,\chi_2) C_{L }^{ij}(\chi_1,\chi_2) \right]  \,.
\end{align}
As before, the absence of cross-terms is due to $\Delta g$ having a vanishing monopole. The first term is the usual expression, while the second is a multiplicative bias. Limber approximating following appendix~\ref{sec:evaluating_integrals}, we find
\begin{align}\label{eqn:cosmic_shear_limber_integral}
    T_{\ell}^{ij} \approx  & \int \frac{\dchi}{\chi^2} \left[ \bar{g}^{i}(\chi) \bar{g}^{j}(\chi)  P_{ij}\left(\frac{\ell+1/2}{\chi}; z\right) +  \sum_{L } M^{\dg^i \dg^j}_{\ell  L} (\chi,\chi) P_{ij}\left(\frac{L+1/2}{\chi}; z\right)\right]\,.
\end{align}
In section \S\ref{sec:kernel_implications}, we explain that this new term is expected to be positive for both auto- and cross-correlation of tomographic cosmic shear measurements. On scales smaller than those on which there is significant projection anisotropy, the second term above can be approximated with exquisite accuracy by equation~\eqref{eqn:simplified_mode_coupling_bias}, just replacing $\phi$ with $g$ (see, e.g., figure~\ref{fig:shear_bias_shape}).

\subsection{Galaxy-galaxy lensing}

Similarly, cross-correlations of the cosmic shear signal with a sample of lens galaxies are susceptible only to the multiplicative, mode-coupling bias. Given the machinery we have developed, it is easy to show that variations in the redshift distributions of lens and source galaxy samples lead to a measured angular spectrum of the form
\begin{align}
    T_{\ell}^{\kappa g} =&  \frac{1}{2\ell+1} \sum_m \langle \kappa_{\ell m} \delta^{(2D),*}_{\ell m} \rangle \nonumber\\
    = & \int d\chi_1\,d\chi_2 \left[ \bar{g}(\chi_1)\phibar(\chi_2) C_\ell(\chi_1,\chi_2) + \sum_{L } M^{\dg \dphi}_{\ell L}(\chi_1,\chi_2) C_{L }(\chi_1,\chi_2) \right] \,,
\end{align}
where $M^{\dg \dphi}_{\ell L}$ is defined by replacing $\{\dphi^{a}, \dphi^{b}\}\rightarrow \{\Delta g, \dphi\}$ in equations~\eqref{eqn:mode_coupling_matrix} and~\eqref{eqn:cl_dphi_of_r}. Likewise, if projection anisotropy is confined to large angular scales, the mode-coupling term can be approximated on smaller scales by substituting $\{\phi^{a}, \phi^{b}\}\rightarrow \{ g, \phi\}$ in equation~\eqref{eqn:simplified_mode_coupling_bias}. As we will now see, the magnitude and sign of this new contribution depends subtly on the relative arrangement of lens and source galaxy distributions.

\subsection{Implications of the kernels}\label{sec:kernel_implications}
\begin{figure}[t]
    \centering
    \includegraphics[scale=0.715]{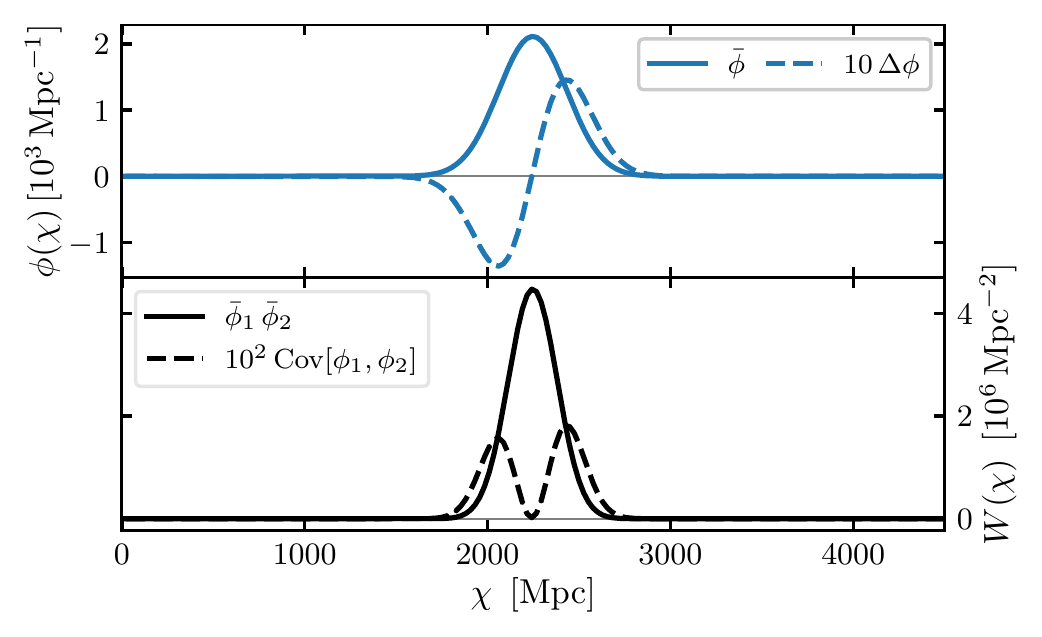}
    \includegraphics[scale=0.715]{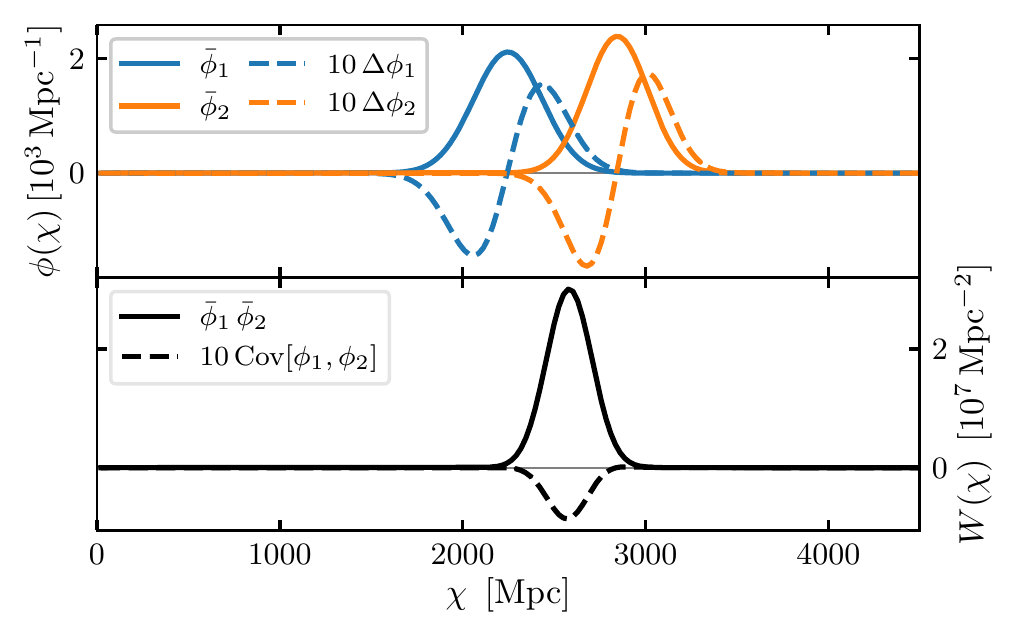}
    \includegraphics[scale=0.715]{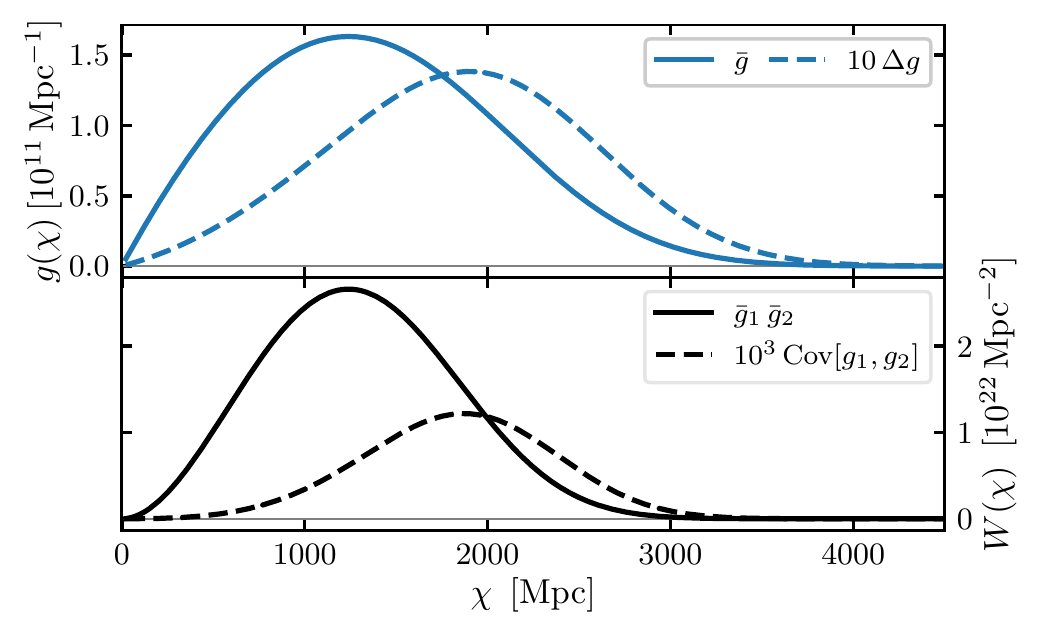}
    \includegraphics[scale=0.715]{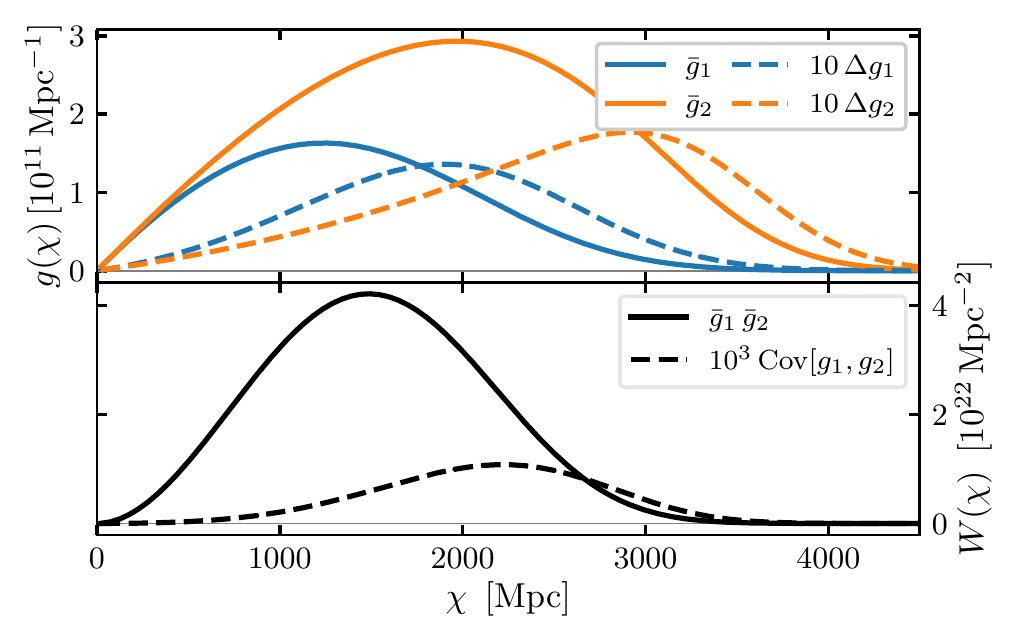}
    \includegraphics[scale=0.715]{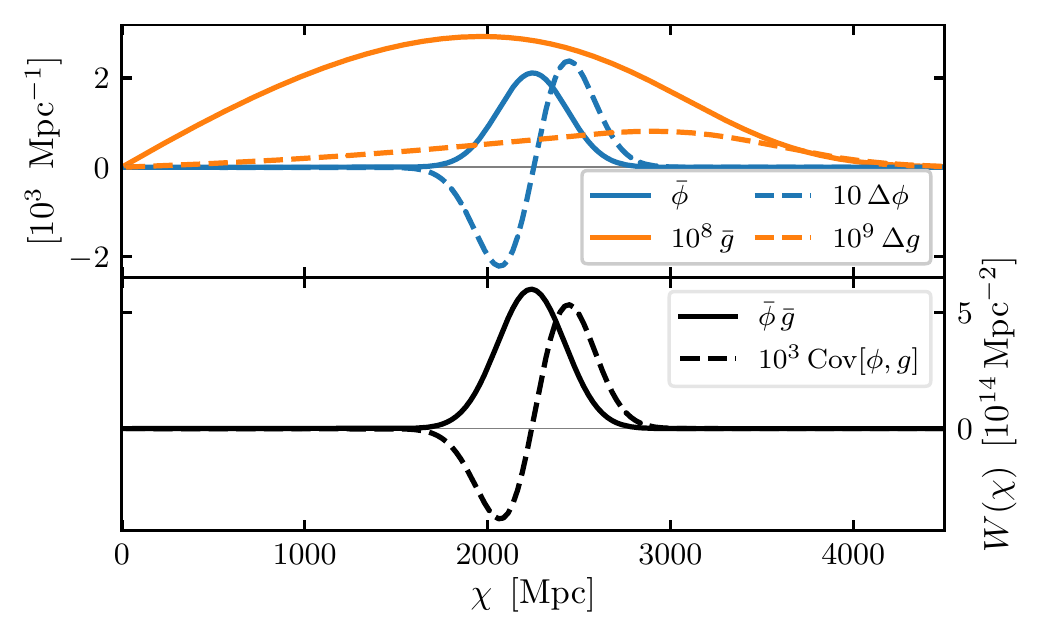} 
    \includegraphics[scale=0.715]{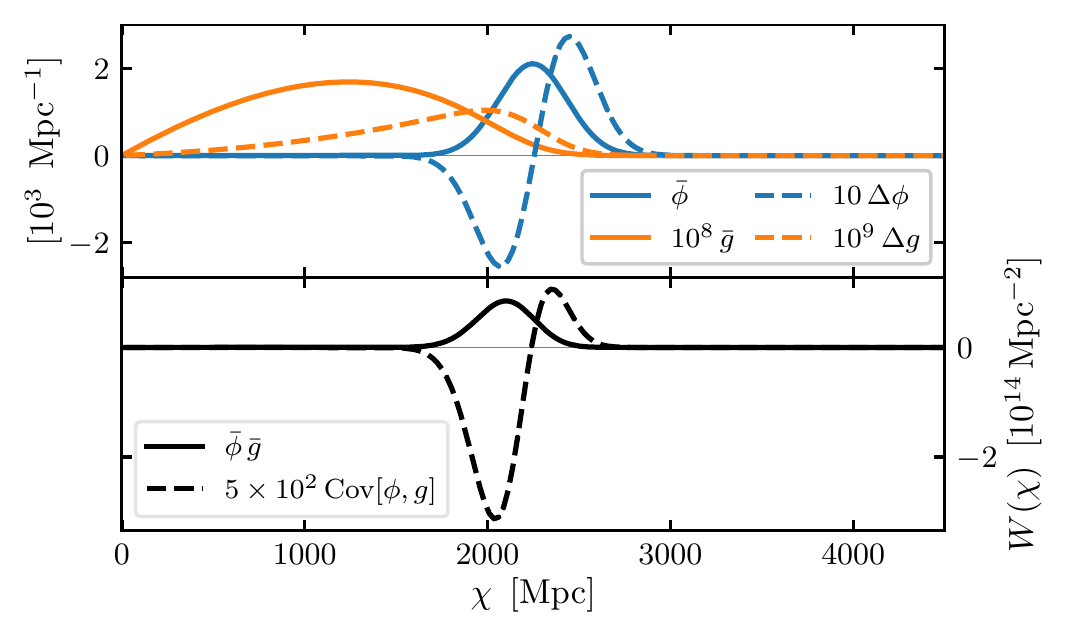} 
    \caption{The signal and mode-coupling kernels for certain special scenarios discussed in \S\ref{sec:special}. These include galaxy clustering auto- (top left) and cross-correlations (top right), the auto- (middle left) and cross-correlation (middle right) of cosmic shear tomographic bins, and galaxy-galaxy lensing analyses where the source galaxies are far behind (bottom left) or close to (bottom right) the lenses.  Within each panel, the top plot shows the unperturbed selection function (solid) along with a random perturbation to it (dashed) associated with a shift of the distribution to higher redshift. The bottom panel then shows the signal kernel (black, solid) and an approximation to the mode-coupling kernel (black, dashed).}
    \label{fig:kernels}
\end{figure}
The kernels for the signal and the mode-coupling term for the special cases above are illustrated in figure \ref{fig:kernels}, from which we can understand much of the resulting phenomenology.  To shift the central redshift of a galaxy bin, $\Delta \phi$ must take the form of a wiggle.  The auto-covariance of this wiggle is a function with two peaks on either side of the peak of $\bar{\phi}$.  We showed in figure \ref{fig:ClDphiChi1Chi2_toymodel} that this structure was an excellent approximation to the full result and led to a bias that had the same shape as the signal.  On the other hand, a shift of the source galaxy distribution to higher/lower redshift is associated with a $\Delta g$ that is consistently positive/negative across $\chi$. 

This has interesting implications if the redshift anisotropy is correlated across both legs of the two-point function: while the mode-coupling contribution to galaxy clustering auto-spectra (figure \ref{fig:kernels}; left, top) and cosmic shear auto- (left, middle) and cross-spectra (right, middle) are always positive, the contribution to the cross-spectrum of different galaxy density bins (right, top) is expected to be negative. The case of galaxy-galaxy lensing is more nuanced: when the source galaxies are far behind the lenses (left, bottom), the mode-coupling kernel is a wiggle that should produce negligible contributions; however, when source and lens galaxies are sufficiently close to each other (right, bottom), $\Delta g$ overlaps only with the lower-$\chi$ part of the $\Delta \phi$ wiggle, and since the two are of opposite sign, the mode-coupling kernel is primarily negative and can potentially lead to significant bias. We shall illustrate these general points with some numerical examples in the following section.

\section{Examples}
\label{sec:examples}

It is worth quantifying the effects we have been describing in the context of some concrete examples. To do so, we will come up with a fiducial redshift distribution and perturb it. We take the galaxy redshift distribution to be Gaussian in comoving distance, albeit a different one in every pixel of a \textsc{Healpix} \cite{ref:healpix_paper} pixelization:
\begin{equation}\label{eqn:pert_dNdz}
    \dNdz(\chi, \nhat) = \frac{1}{ \sqrt{2\pi}\,\sigma_0} e^{-[\chi-\chi'_0(\nhat)]^2 / 2\sigma_0^2}
    \quad , \quad
    \chi'_0(\nhat) \equiv \chi_0 + \chi_{\mathrm{shift}}(\nhat)\,,
\end{equation}
with  $\chi_0$ and $\sigma_0$ some fiducial central distance and standard deviation of the distributions appropriate for the sample at hand, and $\chi_{\mathrm{shift}}(\nhat)$ a random perturbation to the mean.

We generate \textsc{Healpix} \texttt{nside}=128 templates of these perturbations by drawing harmonic coefficients from independent, zero-mean Gaussian distributions with power-law angular power spectra, $C_\ell \propto \ell^{\alpha}$, up to some cutoff scale $\ell'_{\mathrm{max}}$, and adjusting the normalization so that the resulting template map has the desired variance (the quantity that can more easily be quantified observationally). Note that our code, \textsc{CARDiAC}\footnote{Code for Anisotropic Redshift Distributions in Angular Clustering: \url{https://github.com/abaleato/CARDiAC.}}, can also take in user-defined templates of $\chi_{\mathrm{shift}}(\nhat)$ and/or a shift in the width [$\sigma_{\mathrm{shift}}(\nhat)$] and from them calculate the expected contributions.

We obtain the fiducial redshift distribution as an average of~\eqref{eqn:pert_dNdz} over the footprint, construct the fiducial selection function (after assuming a cosmology) and normalize it to satisfy $\int\bar{\phi}\,d\chi=1$. We then use this same normalization to obtain $dn_g/dz$ from the $dN_g/dz$ in equation~\eqref{eqn:pert_dNdz} and thus obtain a spatially-varying selection function as
\begin{equation}\label{eqn:phibar_def}
    \phi(\chi, \nhat) \equiv \frac{H(\chi)}{c} \dndz (\chi, \nhat)\,.
\end{equation}
Finally, we isolate the perturbation as $\dphi(\chi, \nhat) = \phi(\chi, \nhat) - \phibar(\chi)$.

Note that $\phi(\chi, \nhat)$ still respects the integral constraint
\begin{align}
    \int   \dchi \int_{\mathcal{A}} \dnhat \phi(\chi, \nhat) / \int_{\mathcal{A}}\dnhat = 1
    \iff & \int  \dchi \int_{\mathcal{A}} \dnhat  \dphi(\chi, \nhat) = 0 \nonumber \\
    \iff &  \sum_{\ell m}\int  \dchi \dphi_{\ell m}(\chi)\int_{\mathcal{A}}\dnhat  Y^{*}_{\ell m}(\nhat) = 0 \,;
\end{align}
as long as
\begin{align}
    \int \dchi \dphi_{00}(\chi) = 0\,,
\end{align}
but this is satisfied trivially because $\dphi_{00}=0$ by construction\footnote{On a cut sky, there are in principle additional conditions on certain other $\dphi_{\ell m}$'s, those with $\ell$ so small that the corresponding $Y_{\ell m}$'s do not average to zero over the footprint. However, for this same reason, they will be indistinguishable from the monopole for all practical purposes, and will therefore get absorbed into $\phibar$.}. This means that, besides the condition on its average over the analysis region, $\dphi$ is unconstrained.

\subsection{Galaxy clustering}
\label{sec:example_gc}

As a somewhat realistic example, let us consider galaxy samples loosely inspired by the Dark Energy Survey's (DES) \textsc{RedMaGiC}  and \textsc{MagLim} selections, which have been used in several cosmological analyses  (e.g., ref.~\cite{ref:des_y3_cosmo}), both for studies of angular clustering and as samples of lens galaxies in galaxy-galaxy lensing, and estimate the new contributions due to projection anisotropy.

Our calculations will involve a model for the galaxy power spectrum. We obtain this from the \textsc{Anzu}\footnote{\url{https://github.com/kokron/anzu}} code, which combines $N$-body simulations of the dark matter component with an analytic treatment of Lagrangian galaxy bias -- we use the best-fit bias values measured by~\cite{ref:kokron_et_al_21} from simulated samples of \textsc{RedMaGiC}-like galaxies (similar to those in~\cite{ref:derose_et_al_19}) at $z=0.59$.  We assume the fiducial $dN/dz$ is a Gaussian centered at this redshift of $z=0.59$, and with a standard deviation of 0.06 in redshift. This resembles the $dN/dz$ of the third redshift bin of both the \textsc{RedMaGiC} and \textsc{MagLim} samples (see, e.g., figure 1 of~\cite{ref:des_y3_cosmo}).  Note that equation~\eqref{eqn:simplified_mode_coupling_bias} suggests that two samples with similar mean redshift distribution and projection anisotropy will see mode-coupling terms with similar shape and amplitude relative to the signal, independent of the exact galaxy power spectrum (and the same holds for the additive term). Hence, we do not need to consider separately a \textsc{MagLim}-specific galaxy power spectrum.

\begin{figure}
    \centering
    \includegraphics[width=\textwidth]{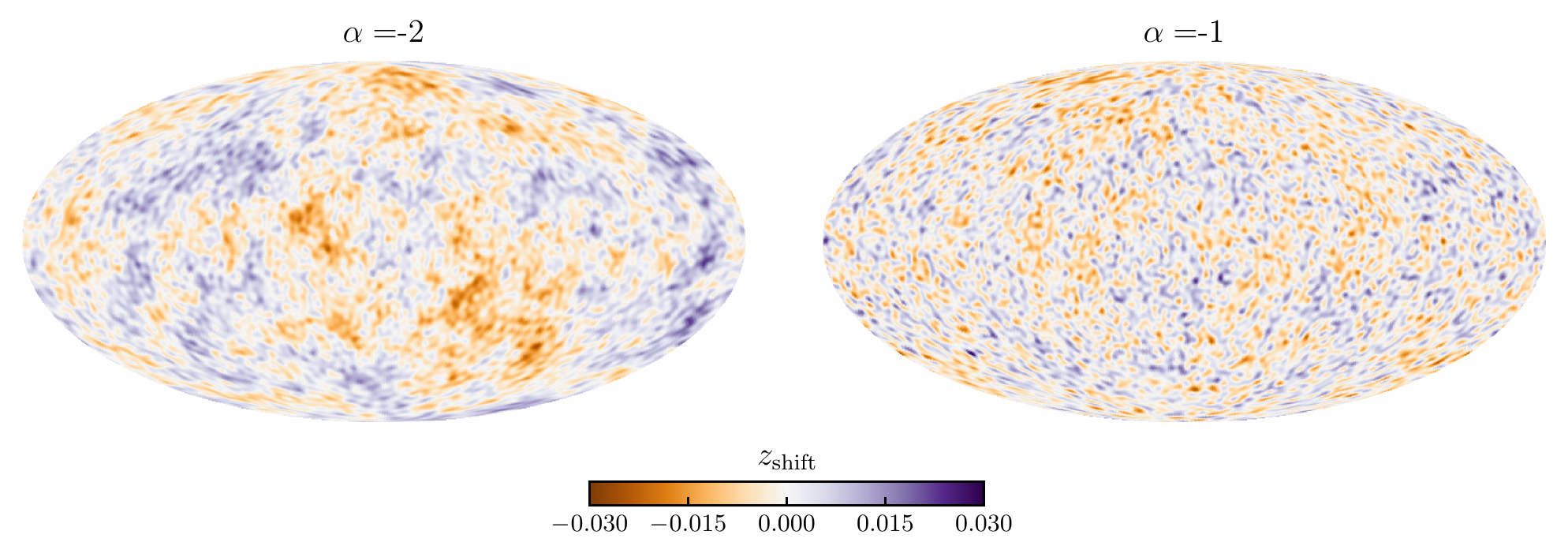}
    \caption{Example variations of $z_{\mathrm{shift}}$ across the sky consistent with the \textsc{MagLim} uncertainty of $\sigma(z_{\mathrm{shift}})=0.006$. We draw these from Gaussian distributions with power-law angular spectra $C_{\ell}\propto \ell^\alpha$, truncated at $\ell_{\mathrm{max}}=100$.
    }
    \label{fig:templates}
\end{figure}

We generate templates of variations in the mean redshift of the $dN/dz$, with characteristics summarized in table~\ref{tab:gc_params}, and two such examples shown in figure~\ref{fig:templates}. As a first approximation, the variance across the shifts template can be related to an uncertainty in the determination of the mean redshift of a given sample (though note that this is likely to underestimate the true variations present in the data). Some values in the literature can give us a sense of the scale of variations appropriate for DES: in their cosmological constraints, the width of their Gaussian prior on the mean redshift of the third bin is $\sigma(z_0)=0.003$ for \textsc{RedMaGiC}, and  $\sigma(z_0)=0.006$ for \textsc{MagLim}~\cite{ref:des_y3_cosmo} both more than an order of magnitude smaller than the bin width. The mean uncertainty on the redshift of a given \textsc{RedMaGiC} galaxy in this bin is much higher, at $\sigma_{z}=0.027$ \cite{ref:elvin-poole_et_al_18}.
\begin{table}[tbp]
    \begin{subtable}[c]{\textwidth}
        \centering
        \begin{tabular}{|c c|c c|}
            \hline
            $z_{0}$ & $\chi_{0}$ [Mpc] & $\sigma_0 \times 10^{2}$ & $\sigma_0$ [Mpc] \\
            \hline
             0.59 & 2241 & $3,\, 6,\, 9,\, 12$  &  $95,\, 187,\, 279,\, 368$  \\
            \hline
        \end{tabular}
    \caption{Parameters characterizing the fiducial $dN/dz$.}
    \end{subtable}
    \newline
    \vspace*{0.2 cm}
    \newline
    \begin{subtable}[c]{\textwidth}
        \centering
        \begin{tabular}{|c|c|c c|}
        \hline
        $\alpha$ & $\ell'_{\mathrm{max}}$ & $\sigma(z_\mathrm{shift})\times 10^{2}$ & $\sigma(\chi_\mathrm{shift})$ [Mpc] \\
        \hline
        $0,\, -1,\, -2$ & 100 &  $0.3,\, 0.6,\, 1.0,\, 2.7$  &  $9.5,\, 19,\, 32,\, 85$  \\
        \hline
        \end{tabular}
    \caption{Parameters characterizing the perturbations.}
    \end{subtable}
    \caption{\label{tab:gc_params} Parameters we consider in our galaxy clustering example. We set $\Delta \phi^{a}=\Delta \phi^{b}$ in order to study the impact on the galaxy clustering power spectrum. Values given in comoving distance units are converted from redshifts assuming the Planck 2018~\cite{ref:planck_18_legacy,ref:planck_params_18} $\Lambda$CDM cosmology.}
\end{table}
\begin{figure}
    \centering
    \includegraphics[width=0.7\textwidth]{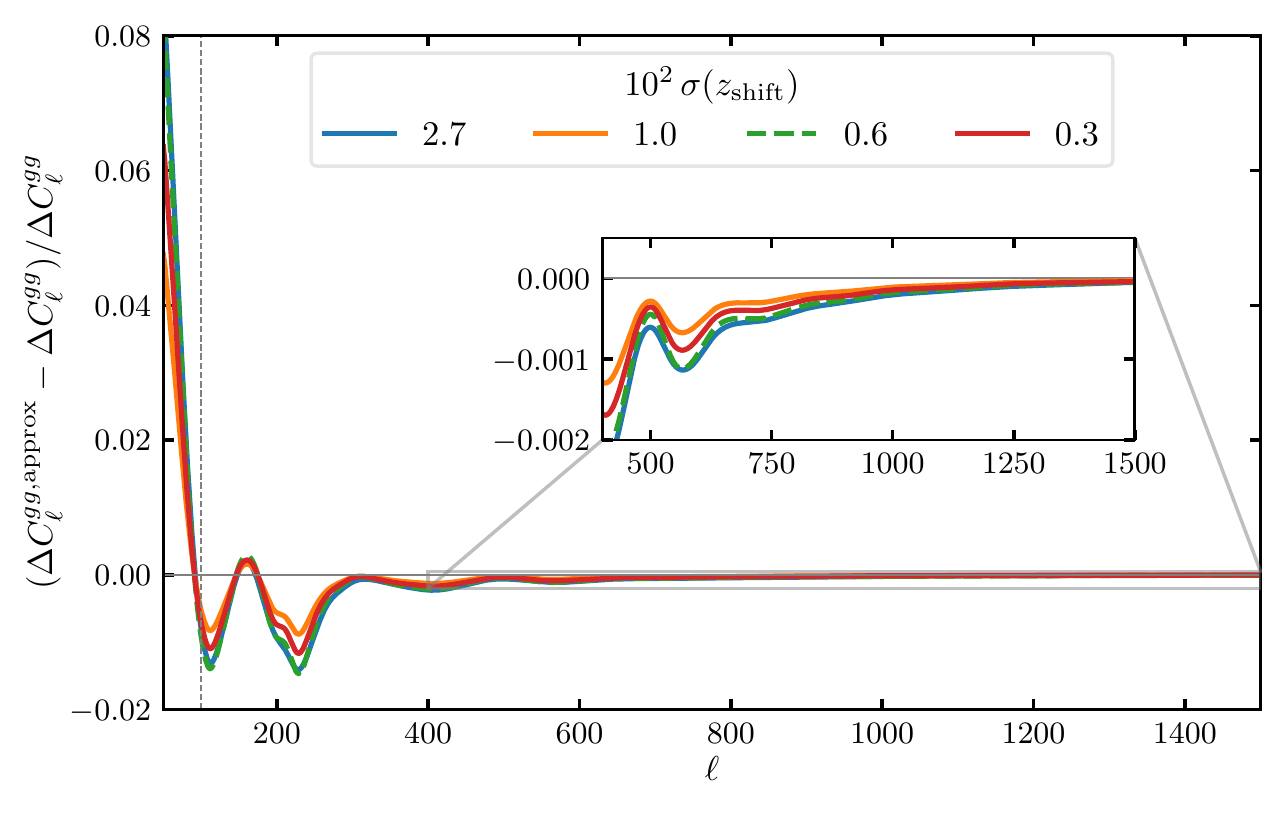}
    \caption{Accuracy of the analytic approximation to the mode-coupling contribution, equation~\eqref{eqn:simplified_mode_coupling_bias}, in the limit where projection anisotropy is restricted to $\ell'<\ell'_{\mathrm{max}}=100$ (dashed, vertical line); in the cases shown, the anisotropy falls off as a power law below the cut, with spectral index $\alpha=-2$ though the result is little changed with $\alpha=-1,0$. The approximation is excellent above scales where there is significant projection anisotropy, for reasons explained in \S\ref{sec:example_gc}.
    }
    \label{fig:analytic_approx}
\end{figure}
\begin{figure}
    \centering
    \includegraphics[width=0.7\textwidth]{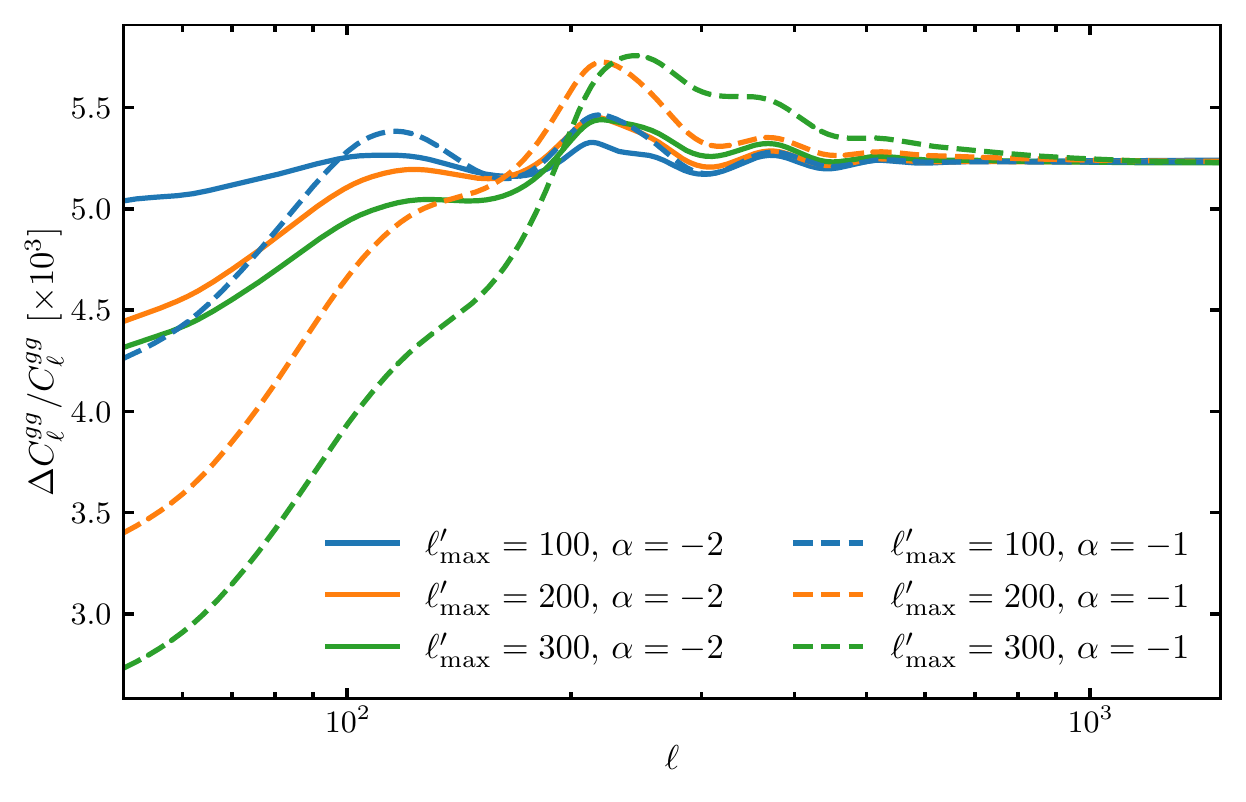}
    \caption{Response of the mode-coupling term to the maximum scale on which there is projection anisotropy. Underlying all curves is the same standard deviation for the mean redshift variations, which we fix to the \textsc{MagLim} value of $\sigma(z_{\mathrm{shift}})=0.6\times 10^{-2}$, though results are qualitatively the same for other values of  $\sigma(z_{\mathrm{shift}})$. The redder the anisotropy spectrum, the faster we converge to the $\ell\gg\ell'$ prediction of flatness from equation~\eqref{eqn:simplified_mode_coupling_bias}.
    }
    \label{fig:comparing_lprime_max}
\end{figure}

We then calculate the mode-coupling contribution in each case working in the Limber approximation; i.e., evaluating equation~\eqref{eqn:modecoupling_bias_in_limber}. We find that the shape of the mode-coupling term tracks the unperturbed signal very closely for $\ell\gg1$ -- i.e., $\Delta C^{gg}_{\ell\gg1}/C^{gg}_{\ell\gg1}$ is flat -- as expected from the discussion around equation~\eqref{eqn:f_toy_model}. Moreover, figure~\ref{fig:analytic_approx} demonstrates that the analytic approximation we developed in equation~\eqref{eqn:simplified_mode_coupling_bias} is in excellent agreement with a full calculation of equation~\eqref{eqn:modecoupling_bias_in_limber}. This is a consequence of the projection anisotropy being confined to large angular scales -- though see figure~\ref{fig:comparing_lprime_max} for the response of the mode-coupling term to changes in the $\ell'_{\mathrm{max}}$ cutoff, which appears to be small.

\begin{figure}
    \centering
    \includegraphics[width=0.7\textwidth]{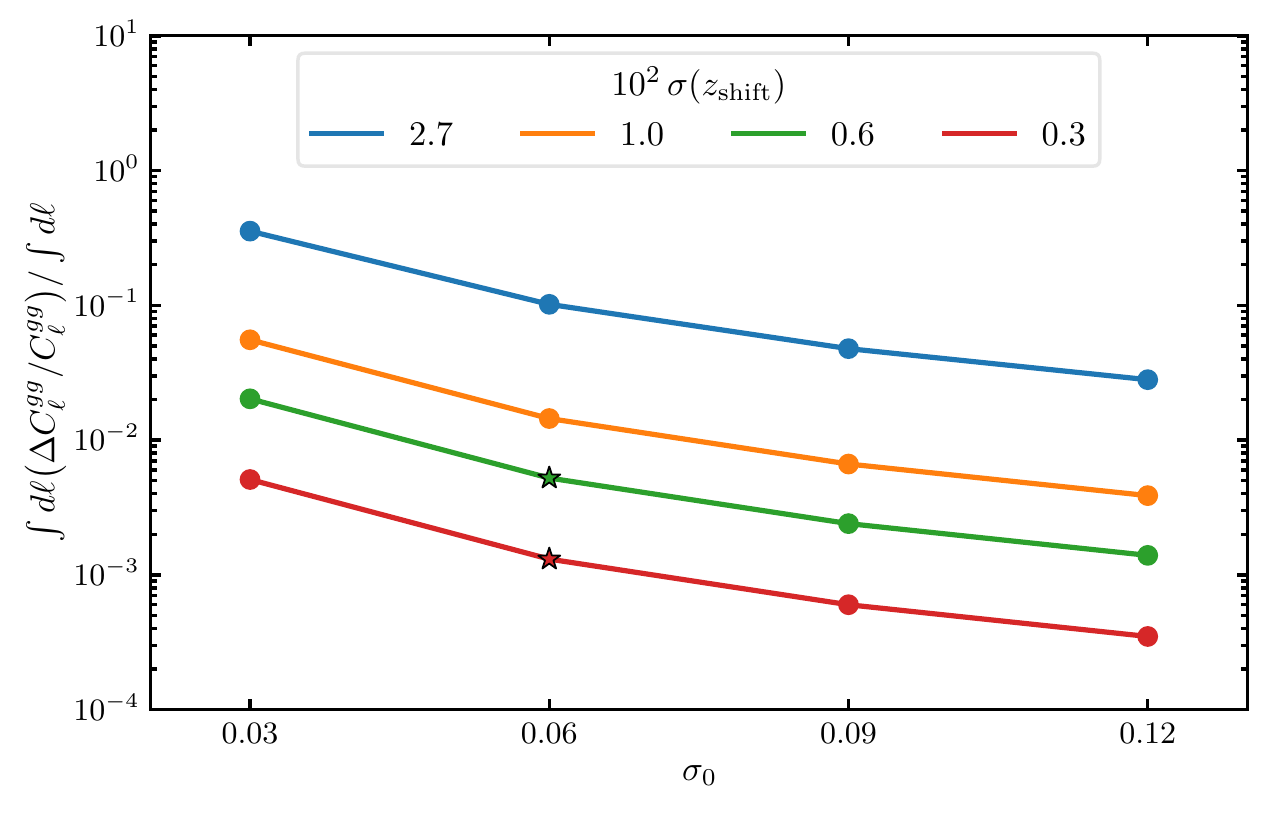}
    \caption{Fractional amplitude of the mode-coupling contribution to the galaxy clustering power spectrum for the samples described in table~\ref{tab:gc_params} and the text. The standard deviation of the injected shifts in central redshift, $\sigma(\chi_{\mathrm{shift}})$, can be compared to uncertainties on the mean redshift of actual samples: the green curve roughly corresponds to \textsc{MagLim} errors, and the red to \textsc{RedMaGiC}. We consider various values for the width of the fiducial distribution, $\sigma_0$, and denote with stars the approximate characteristics of the \textsc{MagLim} and \textsc{RedMaGiC} samples. In the cases shown, the anisotropy follows a power-law with spectral index $\alpha=-2$, truncated at $\ell'_{\mathrm{max}}=100$, though the result is little changed with $\alpha=-1,0$ (see figure~\ref{fig:comparing_lprime_max}).}
    \label{fig:gc_bias_vs_sigma}
\end{figure}

Given the flatness of $\Delta C_{\ell}/C_{\ell}$, we can define a `bias amplitude', 
\begin{equation}\label{eqn:bias_amplitude}
    \int_{\ell=50}^{1500} d\ell \left(\Delta C_{\ell}/C_{\ell}\right) / \int_{\ell=50}^{1500} d\ell\,,
\end{equation}
where the lower end of the range of integration is set by the validity of the Limber approximation. In figure~\ref{fig:gc_bias_vs_sigma} we plot this metric against $\sigma_0$, the width of the fiducial distribution, for various galaxy samples. At fixed $\sigma_0$, the bias amplitude scales as $\Delta C_{\ell} / C_{\ell} \propto \sigma^{2}(\chi_{\mathrm{shift}})$, as identified in equation~\eqref{eqn:f_toy_model}. That same equation also predicted the scaling $\Delta C_{\ell} \propto \sigma^{-4}_0$; however, making the $dN/dz$ narrower also increases the clustering signal. All in all, the bias amplitude goes approximately as
\begin{equation}
    \frac{\Delta C_{\ell}}{C_{\ell}} \propto \left[\frac{\sigma(\chi_{\mathrm{shift}})}{\sigma_0}\right]^2
    \quad .
\label{eqn:bias_scaling}
\end{equation}

The stars in figure~\ref{fig:gc_bias_vs_sigma} denote the rough characteristics of the DES lens galaxy samples. If their redshift uncertainties and distribution width are well characterized, our results suggest that the mode-coupling effect should be negligible for them: a $0.5\%$ and $0.1\%$ correction for the \textsc{MagLim} and \textsc{RedMaGiC} bins we have looked at, respectively. We also calculate the additive contribution and find it to be completely negligible.

So far, we have looked only at the auto-correlation of galaxy overdensity bins. However, in \S\ref{sec:kernel_implications}, we anticipated that the cross-correlation of two samples that are only partially overlapping should be especially affected by mode-coupling biases when the anisotropy is correlated across both samples. To put this on a more quantitative footing, we consider two bins inspired by the DES lens galaxy samples: one centered at $z_0=0.59$, the other at $z_0=0.79$, and both with width $\sigma_0=0.06$. We allow for a level of anisotropy consistent with the \textsc{RedMaGiC} [$\sigma(z_{\mathrm{shift}})= 0.003$] or \textsc{MagLim} errors [$\sigma(z_{\mathrm{shift}})= 0.006$], make the variations common to both fields, and propagate this through to biases on the angular cross-spectrum.

\begin{figure}
    \centering
    \includegraphics[width=0.6\textwidth]{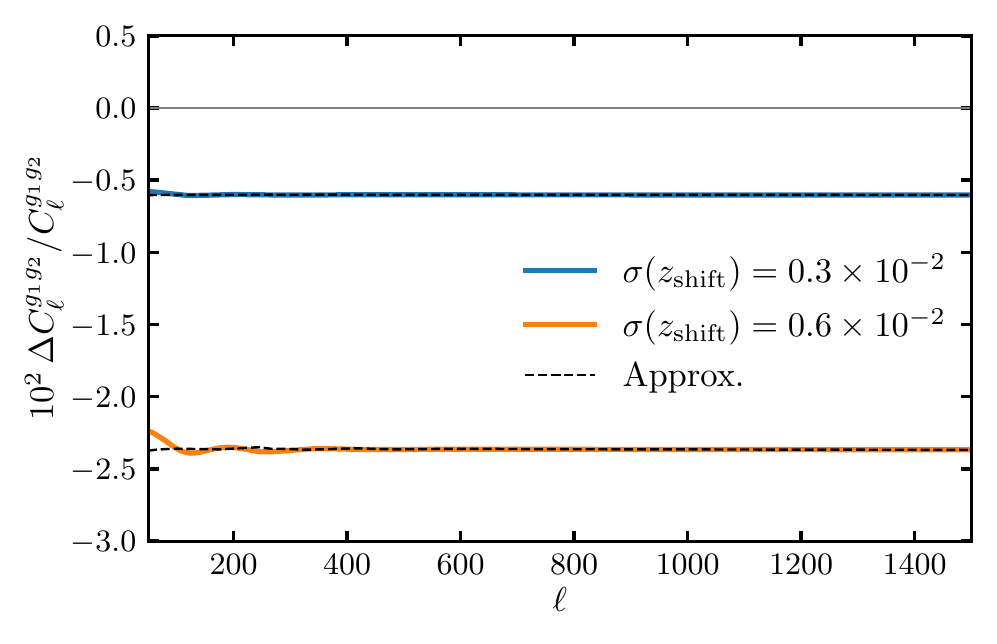}
    \caption{Fractional mode-coupling bias to the cross-correlation of two partially-overlapping galaxy density bins in the limit where the projection anisotropy is common to both bins. The cases we show roughly correspond to cross-correlating two adjacent bins of the \textsc{RedMaGiC} (blue) or \textsc{MagLim} (orange) samples. The dashed lines show an analytic approximation based on the anisotropy covariance, equation~\eqref{eqn:simplified_mode_coupling_bias}. Note the negative sign of this contribution.}
    \label{fig:gg_cross_biases}
\end{figure}

We show our results in figure~\ref{fig:gg_cross_biases}. As expected from the reasoning in \S\ref{sec:kernel_implications} the biases are negative, have very similar shape to the signal, and are very well approximated by the analytic expression in equation~\eqref{eqn:simplified_mode_coupling_bias} based on the anisotropy covariance. This effect could be present at the level of a couple percent for \textsc{MagLim}, or half a precent for \textsc{RedMaGiC}.

\subsection{Cosmic shear}

Next, let us explore a quantitative example in the realm of cosmic shear. For simplicity, we will consider the shear auto-spectrum of source galaxies in a single redshift bin. The more general case of cross-correlations between bins can be studied as required using our publicly-available code, {CARDiAC}. From the discussion in \S\ref{sec:kernel_implications}, we expect no major qualitative difference between the two scenarios.

As in the previous example, we generate templates of $z_{\mathrm{shift}}$ -- a spatially-varying shift in the mean redshift of the source galaxy $dN/dz$ -- with variance $\sigma(z_{\mathrm{shift}})$, for several choices of the width of the fiducial distribution, parametrized by $\sigma_0$.  For this example we consider also various values for the central redshift of the fiducial distribution; for galaxy clustering we did not do this because we expect $\Delta C^{gg}_{\ell}/C^{gg}_{\ell}$ to be independent of $z_0$, all other things being equal. The range of parameter values we explore is detailed in table~\ref{tab:cs_params}. These ranges encompass values that roughly correspond to the DES source galaxy samples summarized in figure~1 and table~1 of ref.~\cite{ref:des_y3_cosmo};  where relevant, we will identify them as such in our plots. The last ingredient we need to evaluate equation~\eqref{eqn:cosmic_shear_limber_integral} is the non-linear matter power spectrum, for which we use the ref.~\cite{ref:mead_et_al_20} version of the HaloFit prescription \cite{ref:smith_et_al_03} as implemented in \textsc{CAMB}~\cite{ref:lewis_challinor_lasenby_99}.
\begin{table}[tbp]
    \begin{subtable}[c]{\textwidth}
        \centering
        \begin{tabular}{|c c|c|}
            \hline
            $z_{0}$ & $\chi_{0}$ [Mpc] & $\sigma_0 \times 10$  \\
            \hline
            $0.2,\, 0.4,\, 0.7,\, 1$ & $844,\, 1600,\, 2579,\, 3396$ & $1,\, 2,\, 3,\, 4$  \\
            \hline
        \end{tabular}
        \caption{Parameters characterizing the fiducial $dN/dz$ of the source galaxies.}
    \end{subtable}
    \newline
    \vspace*{0.2 cm}
    \newline
    \begin{subtable}[c]{\textwidth}
        \centering
        \begin{tabular}{|c|c|c|}
            \hline
            $\alpha$ & $\ell'_{\mathrm{max}}$ & $\sigma(z_\mathrm{shift})\times 10^{2}$ \\
            \hline
            $-2$ & 100 &  $0.1,\, 0.6,\, 1.1,\, 1.5,\, 2.0$ \\
            \hline
        \end{tabular}
        \caption{Parameters characterizing the perturbations.}
    \end{subtable}
    \caption{\label{tab:cs_params} Parameters we consider in our cosmic shear examples. We set $\Delta g^{a}=\Delta g^{b}$ in order to study the impact on the cosmic shear auto-spectrum. Values given in comoving distance units are converted from redshifts assuming the Planck 2018~\cite{ref:planck_18_legacy,ref:planck_params_18} $\Lambda$CDM cosmology. Since the conversion is different at each redshift, we quote the width of the fiducial and perturbation distributions in redshift units only.}
\end{table}

\begin{figure}
    \centering
    \includegraphics[width=0.7\textwidth]{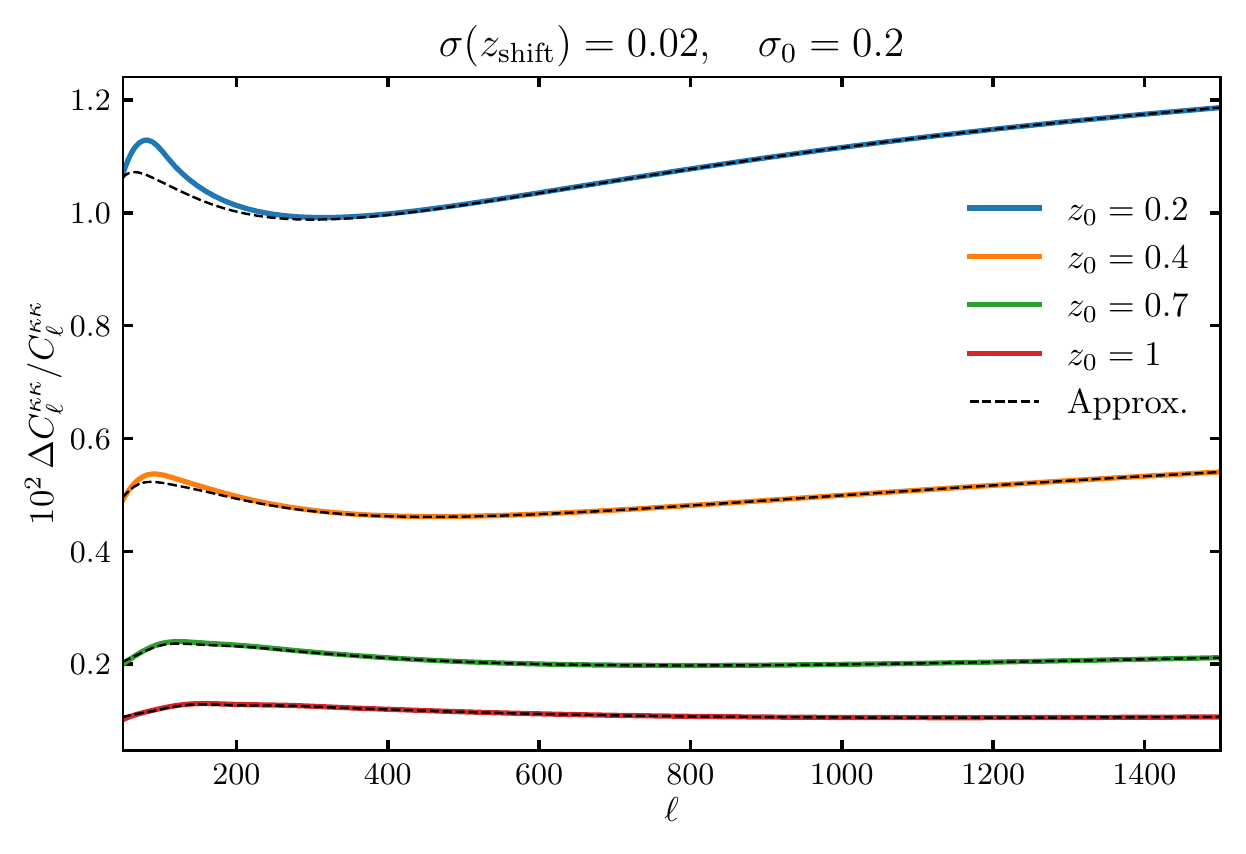}
    \caption{Fractional amplitude of the mode-coupling contribution to the shear power spectrum for a specific example where the source galaxy distribution is centered at $z_0$, has width $\sigma_0=0.2$, and variations about the central redshift with standard deviation $\sigma(z_{\mathrm{shift}})=0.02$ (similar to the DES source galaxy samples). The mode-coupling term has the shape of the signal when the sources are at high redshift, and begins to deviate from it when they are below $z_0\lesssim 0.4$. In either regime, equation~\eqref{eqn:simplified_mode_coupling_bias} provides an excellent approximation (dashed) to the full calculation (solid) at high $\ell$. The behavior seen here extends to other values of $\sigma_0$ and $\sigma(z_{\mathrm{shift}})$.}
    \label{fig:shear_bias_shape}
\end{figure}

Considering once again the limit where the projection anisotropy is confined to large angular scales ($\ell'_{\mathrm{max}}=100$, $\alpha=-2$), we find that the analytic approximation in equation~\eqref{eqn:simplified_mode_coupling_bias} -- replacing $\phi$ with $g$ -- is in exquisite agreement with the full calculation on scales smaller than the injected anisotropy, as evidenced by how accurately the dashed curves (the approximation) track the solid ones (the full calculation) in the example scenario of figure~\ref{fig:shear_bias_shape}. At high enough redshift, $z_0\gtrsim 0.4$, the shape of the mode-coupling term resembles that of the signal. However, when the source galaxies are closer by, the integration kernels of bias and signal begin to differ significantly (see e.g. the right panel of figure~\ref{fig:toy_model_kernels}) and thus $\Delta C^{\kappa \kappa}_{\ell}/C^{\kappa \kappa}_{\ell}$ begins to deviate from flatness, with the departure being greater for narrower source galaxy distributions (lower $\sigma_0$). But even when this is the case, we find our analytic expression to remain an excellent approximation.

\begin{figure}
    \includegraphics[width=0.9\textwidth,right]{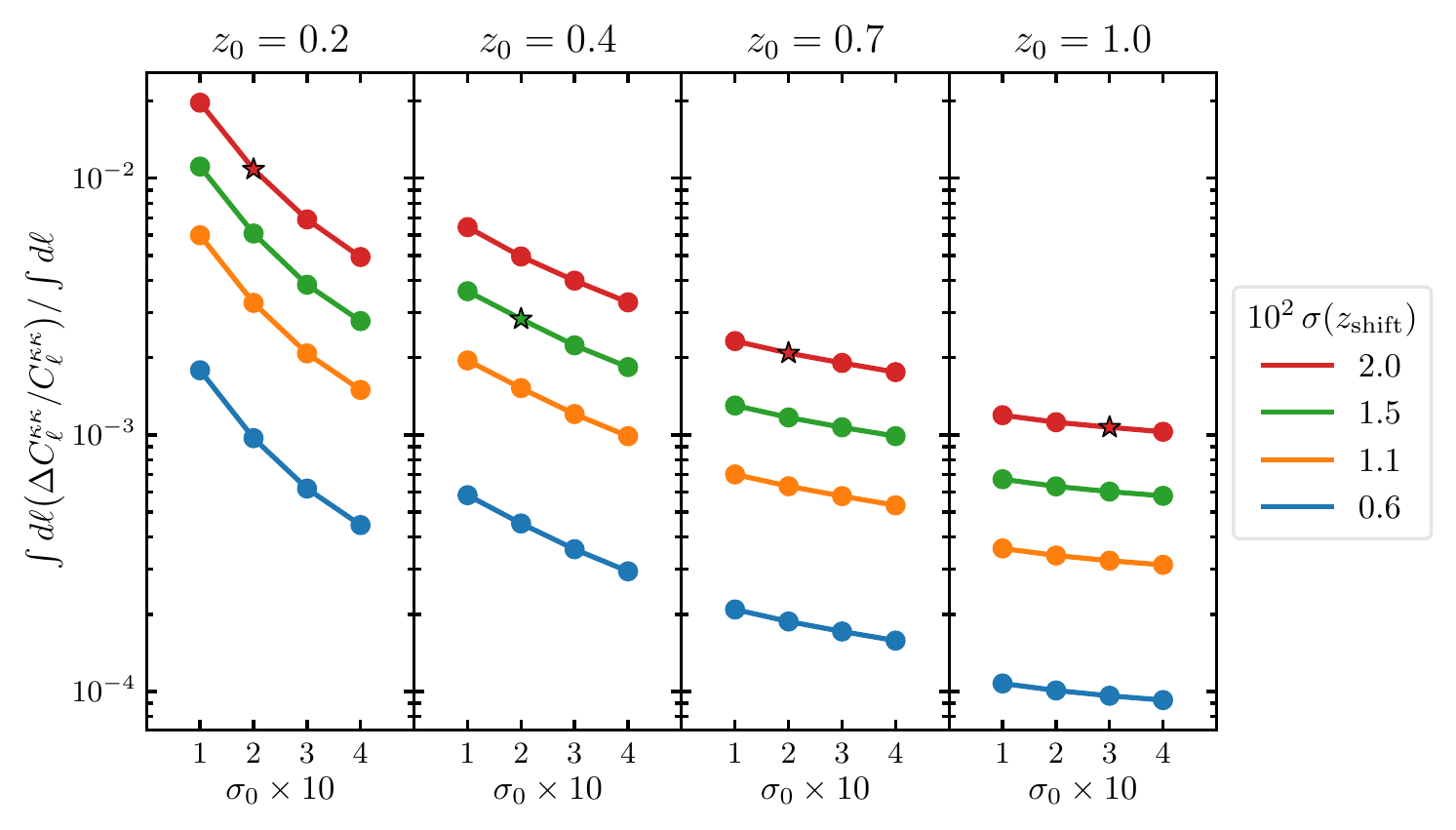}
    \caption{Mean fractional bias to the cosmic shear angular power spectrum for source galaxy redshift bins centered at $z_0$ and with width $\sigma_0$, as a function of $\sigma(z_{\rm shift})$, the scale of variation of the mean redshift across the sky. Stars denote the approximate characteristics of the DES samples.
    }
    \label{fig:gc_bias_vs_sigma_shear}
\end{figure}

Figure \ref{fig:gc_bias_vs_sigma_shear} shows the average fractional bias in $C_\ell^{\kappa\kappa}$ as a function of redshift width ($\sigma_0$) for four different mean redshifts ($z_0$).  Different lines indicate the scale of variation of the mean redshift across the sky, $\sigma(z_{\rm shift})$.  Black stars mark values approximately consistent with those of source galaxy samples in DES, from which we see that the bias is expected to be subdominant to the statistical error except at the lowest redshift bin, where it can be a percent-level effect.

\subsection{Galaxy-galaxy lensing}
Having looked at galaxy clustering and cosmic shear, let us study the intersection of the two: galaxy-galaxy lensing. In particular, let us address the case where there is correlated projection anisotropy across the lens and source galaxy samples. In $\S$~\ref{sec:kernel_implications}, we flagged this scenario as being of special interest because when this is the case and both distributions are close together, the mode-coupling bias is expected to be amplified and negative.

To verify this, we consider a lens galaxy sample with width $\sigma_0^{\mathrm{lens}}=0.06$ centered at $z_0=0.59$, and source samples centered at $z_0^{\mathrm{source}}=0.59$ and $z_0^{\mathrm{source}}=0.8$ with various possible redshift widths. We allow for a level of projection anisotropy in either sample consistent with DES estimates, but ensure that the anisotropy templates in both fields are scaled versions of each other.

\begin{figure}
    \centering 
    \includegraphics[scale=0.75]{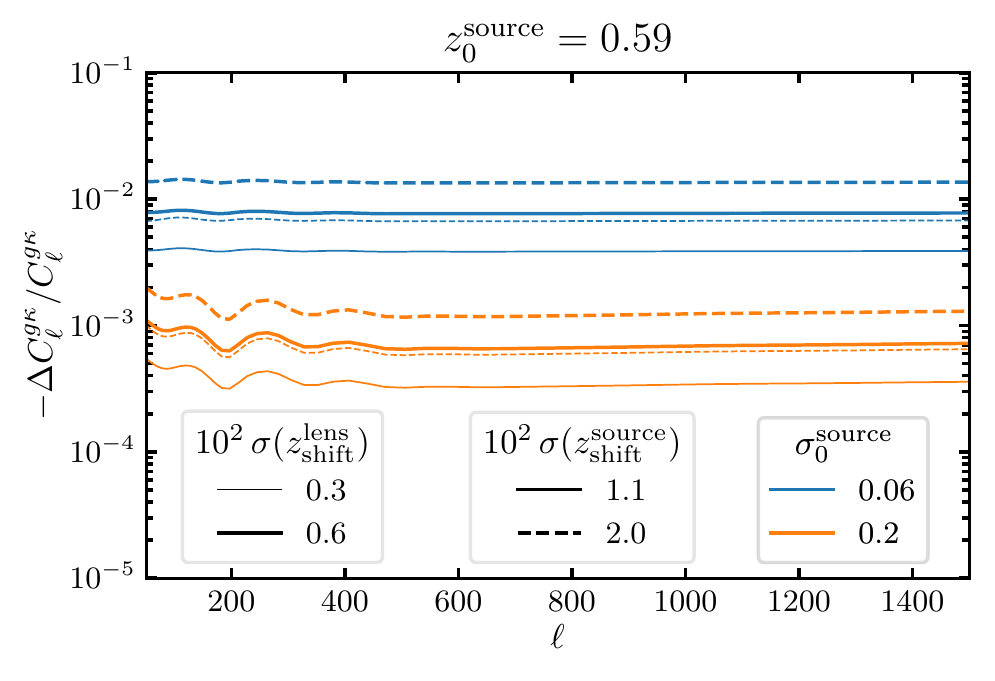}
    \includegraphics[scale=0.75]{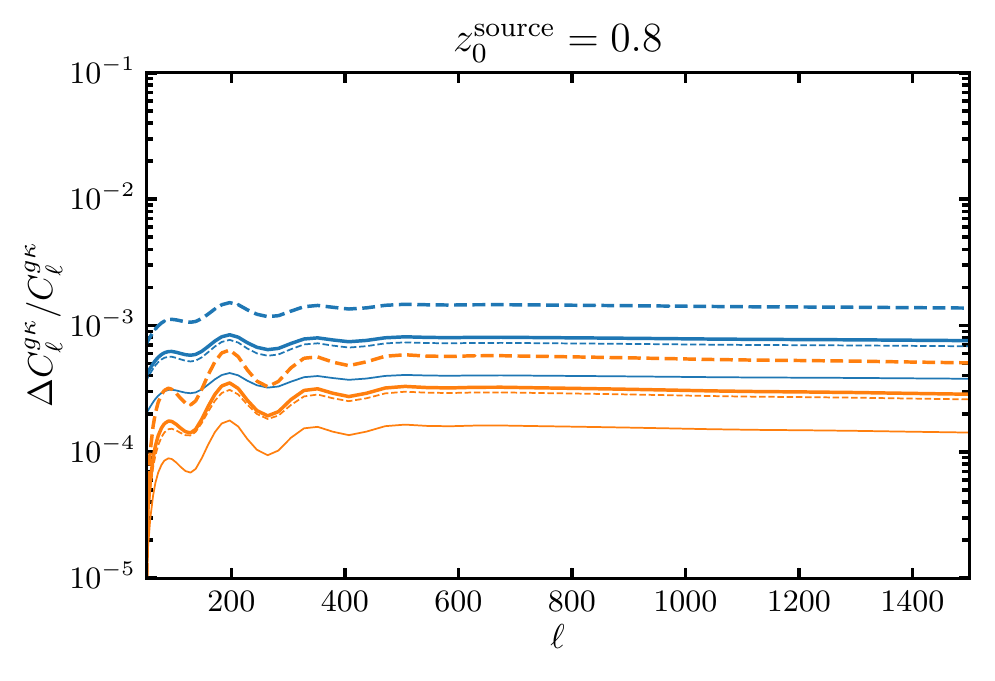} 
    \caption{Fractional mode-coupling bias to the galaxy-galaxy lensing angular spectrum. Without loss of generality, we fix the lens galaxy distribution to have width $\sigma_0^{\mathrm{lens}}=0.06$ and be centered at $z_0^{\mathrm{lens}}=0.59$, and consider two values for each of $\sigma(z_{\mathrm{shift}}^{{\mathrm{lens}}})$, $\sigma(z_{\mathrm{shift}}^{{\mathrm{source}}})$ and $\sigma_0^{{\mathrm{source}}}$, denoting them via line width, line style and color, respectively. In the figure on the right, the sources are far behind the lenses and the biases are all positive; in the one on the left, the two distributions overlap, and the biases are all negative and amplified.}
    \label{fig:ggl_mc_bias}
\end{figure}

The right panel of figure~\ref{fig:ggl_mc_bias} shows that when the sources are far behind the lenses, the bias is positive and small. This can be understood from the bottom left panel of figure~\ref{fig:kernels}: the bias kernel is a wiggle across the integration domain, which leads to extensive cancellations.

On the other hand, when the sources are centered near the lenses, the situation changes dramatically. To understand why, it is useful to keep in mind the bottom right panel of figure~\ref{fig:kernels}. For a narrow enough distribution of sources, $\Delta g$ overlaps only with the low-$\chi$ part of the $\Delta \phi$ wiggle, which has its opposite sign when the projection anisotropies in source and lens distributions are positively correlated. Equipped with this intuition, we can understand the left panel of figure~\ref{fig:ggl_mc_bias}, which quantifies the bias in the limiting case where both lenses and sources are centered at the same redshift\footnote{Though we do not show it explicitly in the figures, we verify that in all cases equation~\eqref{eqn:simplified_mode_coupling_bias} provides an excellent approximation to the full calculation.}. Indeed, the biases are all negative, and grow rapidly with decreasing $\sigma_0^{\mathrm{source}}$. The intuition developed around figure~\ref{fig:kernels} tells us that as $\sigma_0^{\mathrm{source}}$ drops, the bias kernel must be converging to a single peak near $\phibar \gbar$, which is why the shapes of signal and bias become more and more similar. All in all, these effects ought to be below percent-level for DES.

\subsection{Multi-modal distributions}
Finally, as a different application of our formalism, let us address the topic of multi-modal redshift distributions.  Interlopers, or catastrophic outliers in photometric redshifts, are salient examples. The presence of such a poorly-characterized component in the distribution may signal challenges in the photometric selection, which in turn suggests that the fraction of galaxies in each of the modes might be uncertain and varying across the footprint.

\begin{figure}
    \centering 
    \includegraphics[scale=0.73]{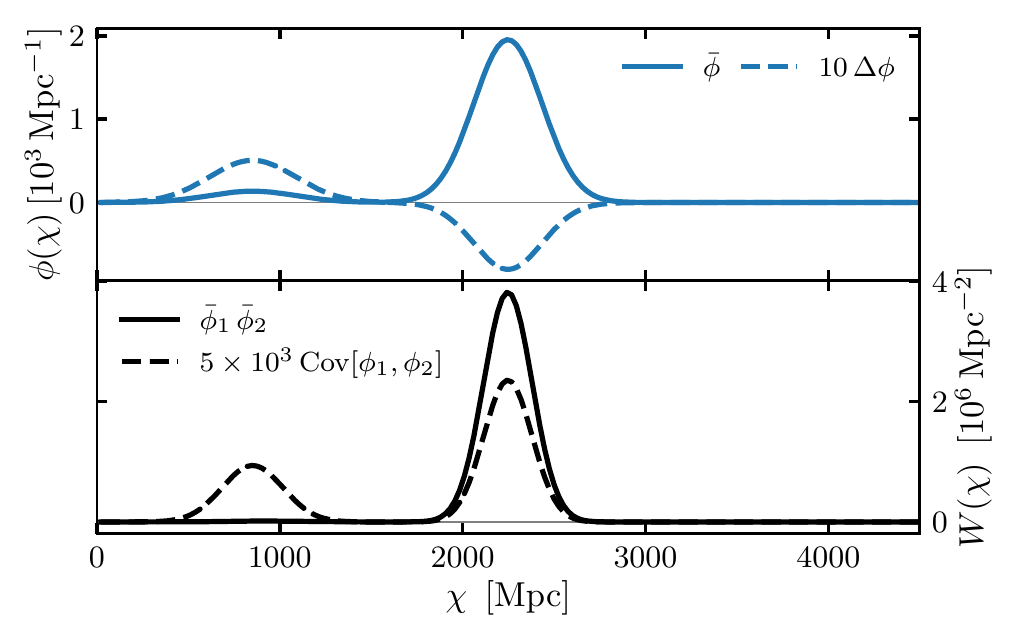}
    \includegraphics[scale=0.73]{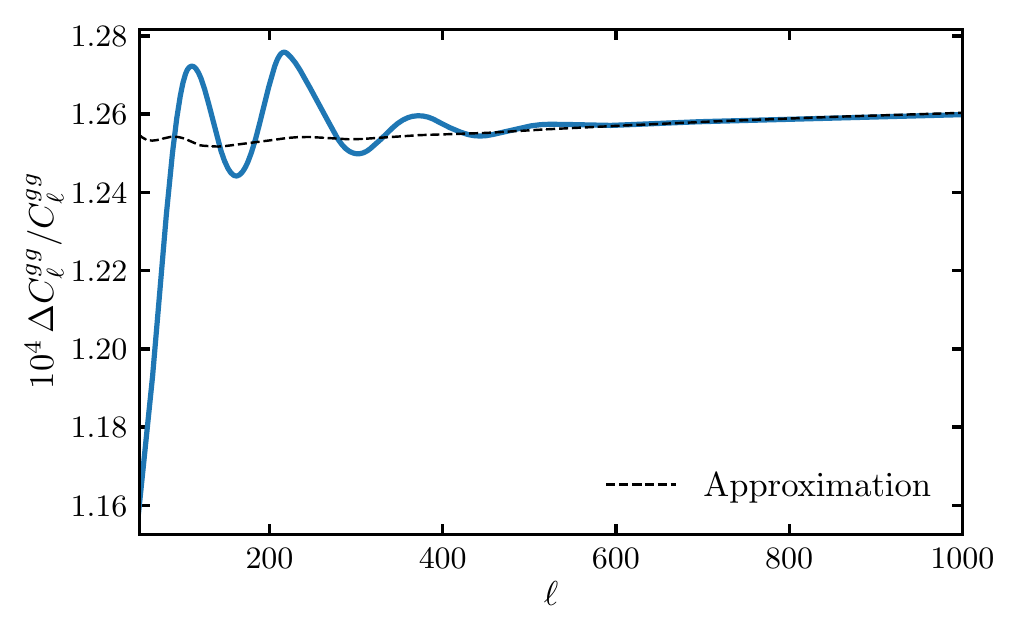} 
    \caption{Impact of spatial variations of a multi-modal $dN/dz$ on the galaxy clustering power spectrum. In the top plot of the left panel, we show the mean distribution (solid) and a typical perturbation to it (dashed); in the bottom plot, we show the signal kernel (solid) along with an approximation to the mode-coupling kernel (dashed). The right panel then shows the associated mode-coupling bias, as calculated using the full expression (solid) or in the approximation of equation~\eqref{eqn:simplified_mode_coupling_bias} (dashed). Because of the interloper component, the $\ell$-to-$k$ mapping is slightly different in the mode-coupling and signal integrals, so $\Delta C^{gg}_{\ell}/C^{gg}_{\ell}$ is not exactly flat.}
    \label{fig:interlopers}
\end{figure}

To be more quantitative, let us suppose, following e.g., \cite{ref:modi_et_al_17}, that the redshift distribution is given by the sum of two components,
\begin{equation}
    \frac{d n_g}{dz} = (1-f) \left(\frac{d n_g}{dz}\right)_{1} + f \left(\frac{d n_g}{dz}\right)_{2}\,,
\end{equation}
where $f$ is the fraction of sources that are misidentified. The top plot in the left panel of figure~\ref{fig:interlopers} shows a plausible such example where the bulk of the mean distribution is at $z=0.59$ and has width $\sigma=0.06$, similar to the DES lens galaxy samples, while $f=0.1$ of the galaxies are interlopers at $z=0.2$ with the same distribution width; we assume both components are Gaussian. We allow $f$ to vary about its mean value across the sky following a red power law ($\alpha=-2$, $\ell'_{\mathrm{max}}=100$) to produce a map-level standard deviation $\sigma(f)=0.01$.

Once again, equation~\eqref{eqn:simplified_mode_coupling_bias} will provide an excellent approximation to the mode-coupling contribution on angular scales smaller than those on which $f$ varies significantly. In this limit, we can use the simplified kernel in the bottom plot of the left panel of figure~\ref{fig:interlopers} for insight. Since the bump at low-$\chi$ induces a different $\ell$-to-$k$ mapping than the signal kernel does, the bias kernel will no longer be as good an approximation to the signal kernel as it was in the discussion around equation~\eqref{eqn:f_toy_model}, and we now expect the shape of the mode-coupling and signal spectra to differ. Indeed, in the right panel of the figure, we see that the bias has acquired a slight blue tilt. Nevertheless, the analytic approximation remains very accurate. As before, the additive contribution appears to be negligible.

These insights carry over to anisotropy in multi-modal redshift distributions of source galaxy samples, and thus have implications for cosmic shear and galaxy-galaxy lensing. However, we defer a rigorous treatment of all possible cases to the practitioners, noting that much of the infrastructure required for the task is present in our publicly-available code.

\section{Conclusions}
\label{sec:conclusions}

The angular auto- and cross-power spectra or correlation functions of fields that are projected onto the sky form one of the key observables from which we extract cosmological information.  As long as the projection kernel, $\phi$, is known, the assumption of statistical isotropy allows us to infer the 3D clustering from the observed, projected clustering.  However, variations in photometry and observational non-idealities across a survey inevitably mean that the projection kernel is anisotropic.  We have shown (\S\ref{sec:formalism}) that such anisotropy leads, in general, to two additional contributions to the observed clustering: an additive term from the auto-correlation of $\Delta\phi$ and a mode-coupling term that arises from the interaction of $\Delta\phi$ with $\delta^{(3D)}$ and couples power at an observed angular wavenumber, $\ell$, to that of nearby scales.

The signal and the two bias terms arise from auto-correlations of the three contributions to the projected density.  This implies that in the special case of a cross-correlation with one field where $\phi$ is `perfectly' known (e.g.\ CMB lensing) both bias contributions vanish.  Since the sky-average of the perturbation to $\phi$ vanishes by definition, if at least one of the fields has mean zero -- as is the case with cosmic shear or galaxy-galaxy lensing -- then only the mode-coupling term survives.  In the general case both can contribute, though the mode-coupling term frequently dominates.

In the limit that the variation, $\Delta\phi$, is primarily on scales much larger than those being probed by the clustering, the mode-coupling contribution to the  angular spectrum is a projection of Cov$[\Delta\phi,\Delta\phi]$ with a known kernel (equation \ref{eqn:simplified_mode_coupling_bias}).  This represents our key result, and allows a simple and accurate estimation of the impact of varying projection across the sky in terms of the observable variance in e.g.\ mean redshift.  Figure \ref{fig:kernels} illustrates the key ingredients for translating the spatial variation in the kernel into the power bias, and can be used to understand multiple special cases described in the text.

We present the implications of our general formalism to several special cases -- including galaxy auto- and cross-correlations, cosmic shear, galaxy-galaxy lensing and CMB lensing -- in \S\ref{sec:special}.  Numerical examples for the impact of shifts in the mean redshift across the sky that follow a power-law power spectrum are presented in \S\ref{sec:examples}, along with a brief exploration of spatially-varying, multi-modal redshift distributions.

We show that for galaxy clustering the mode-coupling bias has a similar shape to the cosmological signal.  For variations in the mean redshift with amplitude $\epsilon$ the bias to the power spectrum is positive and scales as $\epsilon^2/\sigma_0^4$ with $\sigma_0$ the width of the sky-average $dN/dz$.  Since on small scales the clustering signal scales as $\sigma_0^{-2}$ the ratio of bias to signal scales as $(\epsilon/\sigma_0)^2$ (equation \ref{eqn:bias_scaling}).  For current-generation surveys, such as DES, this bias on the power spectrum is at worst percent level\footnote{In our comparisons to current data, we focus on anisotropy in the mean redshift of some $dN/dz$, and assume the extent of the variations is comparable to the uncertainty on the mean redshift quoted by DES. This might be overly optimistic, in which case our results should be regarded as a lower bound on the actual biases.}. In contrast, cross-correlations of galaxy samples can be prone to a negative bias when projection anisotropy is correlated across both tracers; this can also be a percent-level effect for current surveys. 

A similar story holds for cosmic shear (figures \ref{fig:shear_bias_shape} and \ref{fig:gc_bias_vs_sigma_shear}).  Except for redshift distributions with low mean redshift, the quoted uncertainties in the mean redshift from DES would lead to sub-percent biases in the shear auto-spectra that are almost the same shape as the signal. (For the lowest-redshift bin of DES, the bias could be around percent-level.) Such biases would be subdominant to the statistical errors.

The same conclusion holds for galaxy-galaxy lensing (figure~\ref{fig:ggl_mc_bias}), though this case is interesting because the amplitude and sign of the bias depends on the distance between lens and source galaxy distributions -- the bias is amplified when the two are close together, in which case it takes a negative sign. Nevertheless, it is a negligible effect for current surveys.

In closing we have presented a general formalism that allows one to assess the bias introduced on angular clustering measurements of 2D fields by anisotropic projection kernels. We make available a code, \textsc{CARDiAC}\footnote{Code for Anisotropic Redshift Distributions in Angular Clustering: \url{https://github.com/abaleato/CARDiAC.}}, that allows the user to do the same for any specific application. While we have illustrated the formalism with multiple examples, we leave a detailed exploration of specific scenarios to the groups analyzing particular observations. We have further assumed that the $dN/dz$, while anisotropic, is perfectly known, in which case our formalism explains how this information can be incorporated into analyses. We defer consideration of anisotropic and uncertain projection kernels to future work.

\acknowledgments
We are grateful to William Coulton, Emmanuel Schaan, Colin Hill, Giulio Fabbian, Simone Ferraro, Minas Karamanis, Xiao Fang and especially Noah Weaverdyck for useful conversations.
M.W.~is supported by the DOE.
This research has made use of NASA's Astrophysics Data System, the arXiv preprint server, the Python programming language and packages \textsc{NumPy, Matplotlib, SciPy, AstroPy}, \textsc{HealPy}~\cite{ref:healpy_paper}, \textsc{Hankl}~\cite{ref:karamanis_beutler_21} and \textsc{FFTLog-and-Beyond}~\cite{ref:fang_et_al_20}.
This research is supported by the Director, Office of Science, Office of High Energy Physics of the U.S. Department of Energy under Contract No. DE-AC02-05CH11231, and by the National Energy Research Scientific Computing Center, a DOE Office of Science User Facility under the same contract. This work was carried out on the territory of xučyun (Huchiun), the ancestral and unceded land of the Chochenyo speaking Ohlone people, the successors of the sovereign Verona Band of Alameda County.

\appendix

\section{Multiplicative bias from a mischaracterized monopole}\label{appendix:linear_terms}
It is possible to conceive of situations where the footprint-average of the selection function's anisotropy does not vanish, i.e., $\Delta \phi_{00}\neq 0$. This will be the case, for example, when the fiducial $dN/dz$ is calibrated on a patch that does not exactly match that on which the analysis is performed\footnote{One potential example of this is a cross-correlation analysis where the tracers are defined on somewhat different footprints. To avoid the bias described in this section, one could use different fiducial selection functions when projecting to the theoretical angular auto- and cross-spectra, ensuring that the fiducials are always representative of the patch where each measurement is made.}, or when photometric redshifts are systematically offset from their true value. In such a scenario, the general cross-correlation in Eq.~\eqref{eqn:total_cls} receives an additional multiplicative contribution so that
\begin{equation}
    T_\ell^{ab} = \int d\chi_1\,d\chi_2\ \left[ U^{ab}_\ell(\chi_1,\chi_2) + A^{ab}_\ell(\chi_1,\chi_2) + R^{ab}_\ell(\chi_1,\chi_2) + Q^{ab}_\ell(\chi_1,\chi_2)\right]\,.
\end{equation}
The new term, $Q^{ab}_\ell(\chi_1,\chi_2)$, comes from contractions of $\{ \dphi \delta\}$ and $\dphi$ across two projected $\delta^{2D}$s. Let us now derive it.

Statistical isotropy of the $\delta$s implies
\begin{align}
    Q^{ab}_\ell(\chi_1,\chi_2) = \frac{C_{\ell}^{ab}(\chi_1, \chi_2)}{2\ell+1} \sum_{\substack{m\\\ell_1 m_1}} (-1)^{m} G^{\ell \ell_1 \ell}_{-m m_1 m} \left[\phibar(\chi_1)\Delta \phi_{\ell_1 m_1}(\chi_2) + \phibar(\chi_2)\Delta \phi_{\ell_1 m_1}(\chi_1) \right] \,.
\end{align}
We now make use of the selection rules for the Wigner-3j symbols comprising the Gaunt integral, specifically $m_1+m_2+m_3=0$, to obtain
\begin{align}
    G^{\ell \ell_1 \ell}_{-m m_1 m} = (2\ell+1)\sqrt{\frac{(2\ell_1+1)}{4\pi}}\begin{pmatrix}\ell&\ell&\ell_1 \\ m&-m&0 \end{pmatrix} \begin{pmatrix}\ell&\ell&\ell_1 \\ 0&0&0 \end{pmatrix} \delta_{m_1 0} \,.
\end{align}
This, together with the identities
\begin{align}
     \sum_{m} (-1)^{m} \begin{pmatrix}\ell&\ell&\ell_1 \\ m&-m&0 \end{pmatrix}  = (-1)^{\ell} \sqrt{2\ell+1}\ \delta_{\ell_1 0}\,,
\end{align}
and
\begin{align}
     \begin{pmatrix}\ell&\ell&0 \\ 0&0&0 \end{pmatrix}  = \frac{(-1)^{\ell}}{\sqrt{2\ell+1}}\,,
\end{align}
allows us to simplify extensively, giving
\begin{equation}\label{eqn:Q_simplified}
    Q_{\ell}^{ab}(\chi_1, \chi_2) = C_{\ell}^{ab}(\chi_1,\chi_2)   \left[\phibara(\chi_1)\frac{\dphib_{0 0}(\chi_2)}{\sqrt{4\pi}} + \phibarb(\chi_2)\frac{\dphia_{0 0}(\chi_1)}{\sqrt{4\pi}} \right] \,.
\end{equation}
Naturally, only the monopole of $\Delta \phi$ can survive after coupling to an isotropic $\delta$.

\section{Spherical Fourier-Bessel decomposition}\label{sec:sFB}

Let us make contact with the spherical Fourier-Bessel decomposition (sFB)~\cite{Lahav94,Fisher94,HT95,Percival04,Pad01,Pratten13,ref:castorina_white_18a, ref:samushia_19, ref:gebhardt_dore_21, ref:castorina_white_18b, ref:passaglia_et_al_17}. In this language, the arbitrary 3D field defined in equation~\eqref{eqn:glm_of_r} as
\begin{align}
    g_{\ell m}(\chi)\equiv & \int \dnhat Y^*_{\ell m}(\nhat) g(\chi, \nhat)\,,
\end{align}
has conjugate
\begin{align}\label{eqn:glm_of_k}
    g_{\ell m}(k) = \sqrt{\frac{2}{\pi}} k \int \dchi \chi^2 j_{\ell}(k\chi) g_{\ell m}\left(\chi\right)\,.
\end{align}

It will be particularly interesting to work with the sFB decomposition of the perturbation field, $\dphi_{\ell m}(k)$. In \S\ref{sec:full}, we defined its $C_\ell$'s as
\begin{align}
    C_{\ell}^{\dphia \dphib}(\chi_1,\chi_2) \equiv \frac{1}{2\ell+1} \sum_m \dphia_{\ell  m}(\chi_1) \Delta \phi^{b,*}_{\ell  m}(\chi_2)\,.
\end{align}
These can be related to the sFB cross-spectrum
\begin{align}\label{eqn:sfb_ps}
    C_{\ell}^{\dphia \dphib}(k_1,k_2) \equiv \frac{1}{2\ell+1} \sum_m \dphia_{\ell m}(k_1) \Delta \phi^{b,*}_{\ell m}(k_2)\,,
\end{align}
via
\begin{align}\label{eqn:cl_sfb_from_r_to_k}
    C_{\ell}^{\dphia \dphib}(\chi_1,\chi_2) = \frac{2}{\pi} \int \dkone \dktwo k_1 k_2 j_{\ell}(k_1 \chi_1) j_{\ell}(k_2 \chi_2) C_{\ell}^{\dphia \dphib}(k_1,k_2) \,.
\end{align}
These expressions can give us valuable insight into the mode-coupling kernels underlying the effects discussed in the main text. Let us rewrite the perturbation as
\begin{align} 
    \dphi_{\ell m}(k) = & \sqrt{\frac{2}{\pi}} k \int \dchi \chi^2 j_{\ell}(k\chi) \int \dnhat Y^{*}_{\ell m}(\nhat) \dphi(\chi, \nhat) \\
    = & \int \dchi  \chi J_{\ell +1/2}(k\chi) \int \dnhat Y^{*}_{\ell m}(\nhat) \dphi(\chi, \nhat)  \,.
\end{align}
where $J_\mu$ the $\mu$-th order Bessel function of the first kind. This way of writing things makes explicit our approach to evaluating these quantities, which harnesses the computational speed of the \textsc{FFTlog} algorithm for Hankel transforms~\cite{ref:hamilton_00}. The procedure is as follows:
\begin{enumerate}
    \item Define some \textsc{Healpix} pixelization of the sky and generate templates of $z_{\mathrm{shift}}(\nhat_i)$ and/or $\sigma_{\mathrm{shift}}(\nhat_i)$ following $\S$\ref{sec:examples}.
    \item Define a grid of values of $\chi_i$ that are distributed uniformly in logarithmic space, as required by \textsc{FFTlog}. At each $\chi_i$, compute $\dphi(\chi_i, \nhat_i)$ from the $z_{\mathrm{shift}}(\nhat_i)$ and $\sigma_{\mathrm{shift}}(\nhat_i)$ we drew in the previous step, and take the spherical harmomic transform of $\dphi(\chi_i, \nhat_i)$ to obtain  $\dphi_{\ell m}(\chi_i)$.
    \item Use the \textsc{FFTlog} algorithm to compute the $(\ell+1/2)$-th order Hankel tranform of every $m$ mode of $\dphi_{\ell m}(\chi_i)$, obtaining $\dphi_{\ell m}(k_i)$.
\end{enumerate}

\begin{figure}
    \centering
    \includegraphics[width=\textwidth]{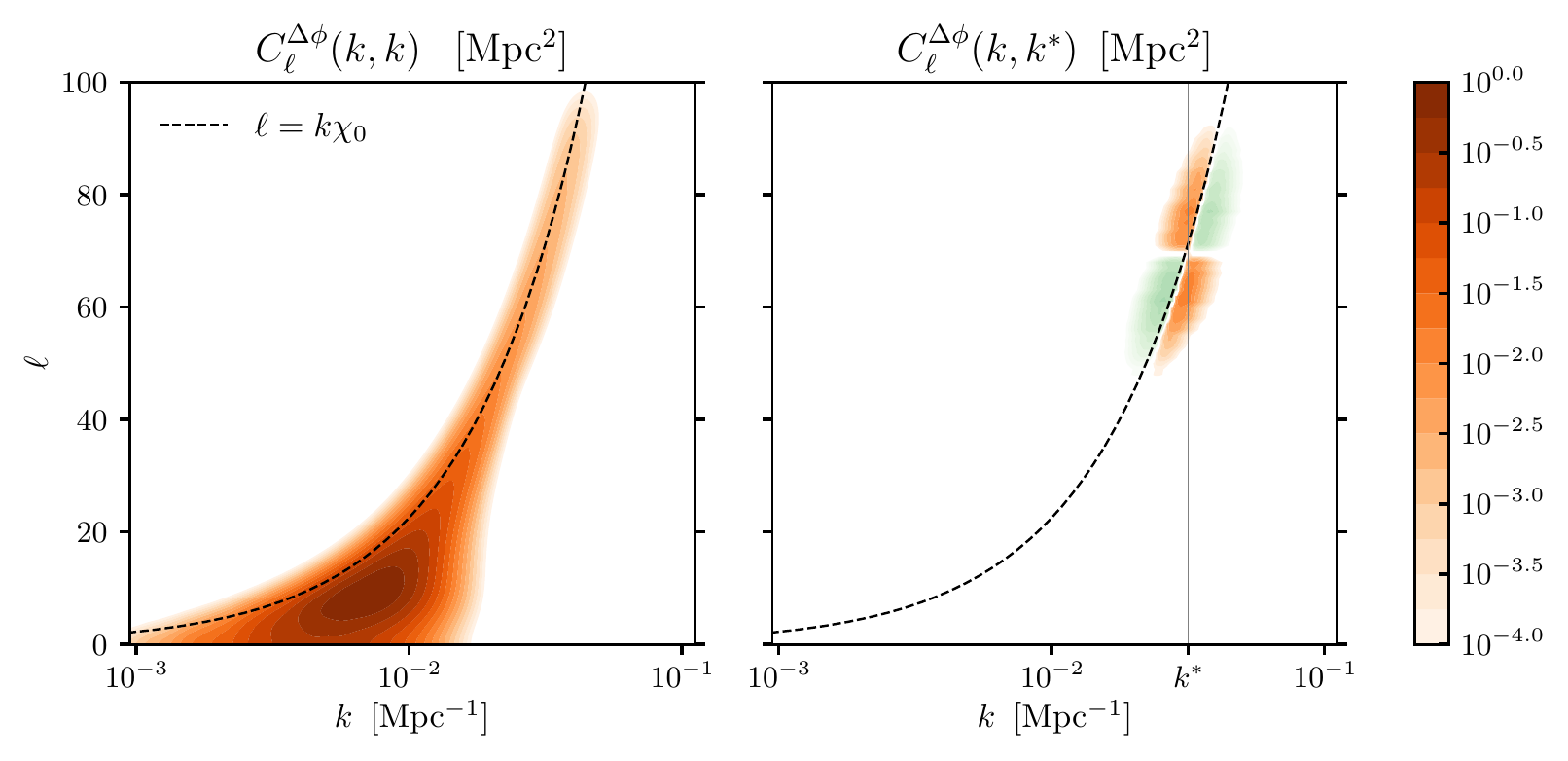}
    \caption{Spherical Fourier-Bessel auto- (left) and cross-spectrum (right, plotted in green where negative). The specific case examined here is produced by the variations in mean redshift shown in the left panel of figure~\ref{fig:templates}, but some key features are general. The left panel shows that the anisotropy power peaks on large angular scales and is sourced primarily by modes transverse to the line-of-sight -- for which $\ell\approx k\chi_0$, where $\chi_0$ is the comoving distance to the center of the distribution -- with this projection being sharper at high $\ell$. This is a common feature of projections and can be related to the structure of the Bessel functions involved, as can be the fact that no anisotropy projects to $\ell \gtrsim k\chi_0$, while some does enter larger scales $\ell < k\chi_0$ via the oscillatory tails of the functions. The right panel shows that off-diagonal spectra are generally weaker, oscillatory, and constrained to a limited region of $k$-space near $k^{*}$.}
    \label{fig:Cldphi_of_k}
\end{figure}

For illustration, we show in figure~\ref{fig:Cldphi_of_k} the sFB auto- and cross-spectrum of an example we have seen previously: the anisotropy produced by the mean-redshift shifts in figure~\ref{fig:templates}.

\section{Evaluating the integrals}\label{sec:evaluating_integrals}

In this section we describe our approach to evaluating the various contributions to equation~\eqref{eqn:total_cls}, the total angular cross-spectrum of two arbitrary fields in light of variations in their redshift distributions.

Consider first the unbiased contribution. Using equation~\eqref{eqn:cls_of_delta} to relate $C^{ab}_\ell(\chi_1,\chi_2)$ to the 3D power spectrum of the tracers, we can write
\begin{align}
    T_{\ell}^{ab} \supset \int d\chi_1\,d\chi_2\ U^{ab}_\ell(\chi_1,\chi_2) =& \int \dchione \dchitwo   \phibara(\chi_1) \phibarb(\chi_2)  \nonumber \\
    & \times \frac{2}{\pi} \int \dk k^2 j_{\ell}(k \chi_1) j_{\ell}(k\chi_2) P_{ab}\left(k; z_1,z_2\right) \,.
\end{align}
In the limit that $P_{ab}$ varies slowly along the radial direction compared to the oscillations of the Bessel functions, it is a good approximation to assume that
\begin{align}\label{eqn:limber_Clab}
     \int \dk k^2 j_{\ell }(k \chi_1) j_{\ell }(k\chi_2) P_{ab}\left(k; z_1,z_2\right) \approx \delta_{\mathrm{D}}(\chi_1 - \chi_2)\ \frac{\pi}{2\chi_1^2} P_{ab}\left(\frac{\ell+1/2}{\chi_1}; z_1\right)\,.
\end{align}
This is the leading-order version of the Limber approximation~\cite{ref:loverde_afshordi_08}. For typical tracers, it works well away from the largest angular scales. In this limit,
\begin{align}\label{eqn:signal_cls}
    T_{\ell}^{ab} \supset \int d\chi_1\,d\chi_2\ U^{ab}_\ell(\chi_1,\chi_2)\approx& \int \dchi \left[ \frac{\phibar(\chi)}{\chi} \right]^2 P_{ab}\left(\frac{\ell+1/2}{\chi}; z\right) \,.
\end{align}
%

Let us now move on to the corrections. We evaluate the additive bias directly as
\begin{align}
    T_{\ell}^{ab} \supset \int d\chi_1\,d\chi_2\ A^{ab}_\ell(\chi_1,\chi_2) = \int d\chi_1\,d\chi_2 \ C_\ell^{\dphia \dphib}(\chi_1,\chi_2) \,,
\end{align}
%
without resorting to Limber. On the other hand, the multiplicative bias 
\begin{align}
    T_{\ell}^{ab} \supset \int d\chi_1\,d\chi_2\ R^{ab}_\ell(\chi_1,\chi_2) =  & \int \dchione \dchitwo   \sum_{L } M^{\dphia \dphib}_{\ell  L} (\chi_1,\chi_2)C_{L }^{ab}(\chi_1,\chi_2)\nonumber\\
    = & \frac{2}{\pi} \int \dchione \dchitwo   \sum_{L } M^{\dphia \dphib}_{\ell  L} (\chi_1,\chi_2) \nonumber \\
    & \times  \int \dk k^2 j_{L }(k \chi_1) j_{L }(k\chi_2) P_{ab}\left(k; z_1,z_2\right) \,,
\end{align}
can be evaluated efficiently by applying Limber to the integral over $k$, independent of the smoothness of the $\dphi$'s, giving
\begin{align}\label{eqn:modecoupling_bias_in_limber}
    T_{\ell}^{ab} \supset \int d\chi_1\,d\chi_2\ R^{ab}_\ell(\chi_1,\chi_2)
    \approx & \int \dchi \frac{1}{\chi^2} \sum_{L } M^{\dphia \dphib}_{\ell  L} (\chi,\chi) P_{ab}\left(\frac{L+1/2}{\chi}; z\right) \,.
\end{align}

\bibliographystyle{JHEP}
\bibliography{main}

\newcommand{\noop}[1]{}

\providecommand{\href}[2]{#2}\begingroup\raggedright\begin{thebibliography}{10}

\bibitem{ref:peebles_80}
P.~J.~E. {Peebles}, \emph{{The large-scale structure of the universe}}. 1980.

\bibitem{ref:peebles_93}
P.~J.~E. {Peebles}, \emph{{Principles of Physical Cosmology}}. 1993.

\bibitem{Peacock99}
J.~A. {Peacock}, \emph{{Cosmological Physics}}. Jan., 1999.

\bibitem{Dodelson03}
S.~{Dodelson}, \emph{{Modern cosmology}}. 2003.

\bibitem{Dodelson20}
S.~Dodelson and F.~Schmidt, \emph{Modern Cosmology}. Elsevier Science, 2020.

\bibitem{Baumann22}
D.~Baumann, \emph{Cosmology}. Cambridge University Press, 2022,
  \href{https://doi.org/10.1017/9781108937092}{10.1017/9781108937092}.

\bibitem{2012arXiv1211.0310L}
{\relax LSST Dark Energy Science Collaboration}, \emph{{Large Synoptic Survey
  Telescope: Dark Energy Science Collaboration}}, {\emph{ArXiv e-prints} (2012)
  } [\href{https://arxiv.org/abs/1211.0310}{{\ttfamily 1211.0310}}].

\bibitem{LSST}
{LSST Science Collaboration}, P.~A. {Abell}, J.~{Allison}, S.~F. {Anderson},
  J.~R. {Andrew}, J.~R.~P. {Angel} et~al., \emph{{LSST Science Book, Version
  2.0}}, {\emph{ArXiv e-prints} (2009) }
  [\href{https://arxiv.org/abs/0912.0201}{{\ttfamily 0912.0201}}].

\bibitem{ref:lsst}
{\v Z}.~{Ivezi{\'c}} et~al., \emph{{LSST: From Science Drivers to Reference
  Design and Anticipated Data Products}},
  \href{https://doi.org/10.3847/1538-4357/ab042c}{\emph{\apj} {\bfseries 873}
  (2019) 111} [\href{https://arxiv.org/abs/0805.2366}{{\ttfamily 0805.2366}}].

\bibitem{ref:euclid_12}
L.~{Amendola} et~al., \emph{{Cosmology and Fundamental Physics with the Euclid
  Satellite}}, \href{https://doi.org/10.12942/lrr-2013-6}{\emph{Living Reviews
  in Relativity} {\bfseries 16} (2013) 6}
  [\href{https://arxiv.org/abs/1206.1225}{{\ttfamily 1206.1225}}].

\bibitem{Amendola18}
L.~{Amendola}, S.~{Appleby}, A.~{Avgoustidis}, D.~{Bacon}, T.~{Baker},
  M.~{Baldi} et~al., \emph{{Cosmology and fundamental physics with the Euclid
  satellite}}, \href{https://doi.org/10.1007/s41114-017-0010-3}{\emph{Living
  Reviews in Relativity} {\bfseries 21} (2018) 2}
  [\href{https://arxiv.org/abs/1606.00180}{{\ttfamily 1606.00180}}].

\bibitem{ref:huterer_et_al_13}
D.~{Huterer}, C.~E. {Cunha} and W.~{Fang}, \emph{{Calibration errors unleashed:
  effects on cosmological parameters and requirements for large-scale structure
  surveys}}, \href{https://doi.org/10.1093/mnras/stt653}{\emph{\mnras}
  {\bfseries 432} (2013) 2945}
  [\href{https://arxiv.org/abs/1211.1015}{{\ttfamily 1211.1015}}].

\bibitem{ref:shafer_and_huterer_15}
D.~L. {Shafer} and D.~{Huterer}, \emph{{Multiplicative errors in the galaxy
  power spectrum: self-calibration of unknown photometric systematics for
  precision cosmology}},
  \href{https://doi.org/10.1093/mnras/stu2640}{\emph{\mnras} {\bfseries 447}
  (2015) 2961} [\href{https://arxiv.org/abs/1410.0035}{{\ttfamily 1410.0035}}].

\bibitem{ref:weaverdyck_and_huterer_20}
N.~{Weaverdyck} and D.~{Huterer}, \emph{{Mitigating contamination in LSS
  surveys: a comparison of methods}},
  \href{https://doi.org/10.1093/mnras/stab709}{\emph{\mnras} {\bfseries 503}
  (2021) 5061} [\href{https://arxiv.org/abs/2007.14499}{{\ttfamily
  2007.14499}}].

\bibitem{Salvato19}
M.~{Salvato}, O.~{Ilbert} and B.~{Hoyle}, \emph{{The many flavours of
  photometric redshifts}},
  \href{https://doi.org/10.1038/s41550-018-0478-0}{\emph{Nature Astronomy}
  {\bfseries 3} (2019) 212} [\href{https://arxiv.org/abs/1805.12574}{{\ttfamily
  1805.12574}}].

\bibitem{Newman22}
J.~A. {Newman} and D.~{Gruen}, \emph{{Photometric Redshifts for Next-Generation
  Surveys}},
  \href{https://doi.org/10.1146/annurev-astro-032122-014611}{\emph{\araa}
  {\bfseries 60} (2022) 363}
  [\href{https://arxiv.org/abs/2206.13633}{{\ttfamily 2206.13633}}].

\bibitem{ref:ouchi_et_al_20}
M.~{Ouchi}, Y.~{Ono} and T.~{Shibuya}, \emph{{Observations of the
  Lyman-{\ensuremath{\alpha}} Universe}},
  \href{https://doi.org/10.1146/annurev-astro-032620-021859}{\emph{\araa}
  {\bfseries 58} (2020) 617}
  [\href{https://arxiv.org/abs/2012.07960}{{\ttfamily 2012.07960}}].

\bibitem{Datta07}
K.~K. {Datta}, T.~R. {Choudhury} and S.~{Bharadwaj}, \emph{{The multifrequency
  angular power spectrum of the epoch of reionization 21-cm signal}},
  \href{https://doi.org/10.1111/j.1365-2966.2007.11747.x}{\emph{\mnras}
  {\bfseries 378} (2007) 119}
  [\href{https://arxiv.org/abs/astro-ph/0605546}{{\ttfamily
  astro-ph/0605546}}].

\bibitem{Shaw14}
J.~R. {Shaw}, K.~{Sigurdson}, U.-L. {Pen}, A.~{Stebbins} and M.~{Sitwell},
  \emph{{All-sky Interferometry with Spherical Harmonic Transit Telescopes}},
  \href{https://doi.org/10.1088/0004-637X/781/2/57}{\emph{\apj} {\bfseries 781}
  (2014) 57} [\href{https://arxiv.org/abs/1302.0327}{{\ttfamily 1302.0327}}].

\bibitem{ref:castorina_white_18a}
E.~{Castorina} and M.~{White}, \emph{{Beyond the plane-parallel approximation
  for redshift surveys}},
  \href{https://doi.org/10.1093/mnras/sty410}{\emph{\mnras} {\bfseries 476}
  (2018) 4403} [\href{https://arxiv.org/abs/1709.09730}{{\ttfamily
  1709.09730}}].

\bibitem{ref:castorina_white_18b}
E.~{Castorina} and M.~{White}, \emph{{The Zeldovich approximation and
  wide-angle redshift-space distortions}},
  \href{https://doi.org/10.1093/mnras/sty1437}{\emph{\mnras} {\bfseries 479}
  (2018) 741} [\href{https://arxiv.org/abs/1803.08185}{{\ttfamily
  1803.08185}}].

\bibitem{ref:hivon_et_al_02}
E.~{Hivon}, K.~M. {G{\'o}rski}, C.~B. {Netterfield}, B.~P. {Crill}, S.~{Prunet}
  and F.~{Hansen}, \emph{{MASTER of the Cosmic Microwave Background Anisotropy
  Power Spectrum: A Fast Method for Statistical Analysis of Large and Complex
  Cosmic Microwave Background Data Sets}},
  \href{https://doi.org/10.1086/338126}{\emph{\apj} {\bfseries 567} (2002) 2}
  [\href{https://arxiv.org/abs/astro-ph/0105302}{{\ttfamily
  astro-ph/0105302}}].

\bibitem{ref:des_y3_cosmo}
{DES Collaboration}, \emph{{Dark Energy Survey Year 3 results: Cosmological
  constraints from galaxy clustering and weak lensing}},
  \href{https://doi.org/10.1103/PhysRevD.105.023520}{\emph{\prd} {\bfseries
  105} (2022) 023520} [\href{https://arxiv.org/abs/2105.13549}{{\ttfamily
  2105.13549}}].

\bibitem{ref:planck_18_legacy}
{Planck Collaboration}, \emph{{Planck 2018 results. I. Overview and the
  cosmological legacy of Planck}}, {\emph{arXiv e-prints} (2018)
  arXiv:1807.06205} [\href{https://arxiv.org/abs/1807.06205}{{\ttfamily
  1807.06205}}].

\bibitem{ref:planck_params_18}
{Planck Collaboration}, \emph{{Planck 2018 results. VI. Cosmological
  parameters}}, \href{https://doi.org/10.1051/0004-6361/201833910}{\emph{\aap}
  {\bfseries 641} (2020) A6}
  [\href{https://arxiv.org/abs/1807.06209}{{\ttfamily 1807.06209}}].

\bibitem{ref:varshalovich_book}
D.~A. {Varshalovich}, A.~N. {Moskalev} and V.~K. {Khersonskii}, \emph{{Quantum
  Theory of Angular Momentum}}. 1988,
  \href{https://doi.org/10.1142/0270}{10.1142/0270}.

\bibitem{ref:kaiser_squires_93}
N.~{Kaiser} and G.~{Squires}, \emph{{Mapping the Dark Matter with Weak
  Gravitational Lensing}}, \href{https://doi.org/10.1086/172297}{\emph{\apj}
  {\bfseries 404} (1993) 441}.

\bibitem{ref:healpix_paper}
K.~M. {G{\'o}rski}, E.~{Hivon}, A.~J. {Banday}, B.~D. {Wandelt}, F.~K.
  {Hansen}, M.~{Reinecke} et~al., \emph{{HEALPix: A Framework for
  High-Resolution Discretization and Fast Analysis of Data Distributed on the
  Sphere}}, \href{https://doi.org/10.1086/427976}{\emph{\apj} {\bfseries 622}
  (2005) 759} [\href{https://arxiv.org/abs/arXiv:astro-ph/0409513}{{\ttfamily
  arXiv:astro-ph/0409513}}].

\bibitem{ref:kokron_et_al_21}
N.~{Kokron}, J.~{DeRose}, S.-F. {Chen}, M.~{White} and R.~H. {Wechsler},
  \emph{{The cosmology dependence of galaxy clustering and lensing from a
  hybrid N-body-perturbation theory model}},
  \href{https://doi.org/10.1093/mnras/stab1358}{\emph{\mnras} {\bfseries 505}
  (2021) 1422} [\href{https://arxiv.org/abs/2101.11014}{{\ttfamily
  2101.11014}}].

\bibitem{ref:derose_et_al_19}
J.~{DeRose} et~al., \emph{{The Buzzard Flock: Dark Energy Survey Synthetic Sky
  Catalogs}}, \href{https://doi.org/10.48550/arXiv.1901.02401}{\emph{arXiv
  e-prints} (2019) arXiv:1901.02401}
  [\href{https://arxiv.org/abs/1901.02401}{{\ttfamily 1901.02401}}].

\bibitem{ref:elvin-poole_et_al_18}
J.~{Elvin-Poole} and {DES Collaboration}, \emph{{Dark Energy Survey year 1
  results: Galaxy clustering for combined probes}},
  \href{https://doi.org/10.1103/PhysRevD.98.042006}{\emph{\prd} {\bfseries 98}
  (2018) 042006} [\href{https://arxiv.org/abs/1708.01536}{{\ttfamily
  1708.01536}}].

\bibitem{ref:mead_et_al_20}
A.~J. {Mead}, S.~{Brieden}, T.~{Tr{\"o}ster} and C.~{Heymans},
  \emph{{HMCODE-2020: improved modelling of non-linear cosmological power
  spectra with baryonic feedback}},
  \href{https://doi.org/10.1093/mnras/stab082}{\emph{\mnras} {\bfseries 502}
  (2021) 1401} [\href{https://arxiv.org/abs/2009.01858}{{\ttfamily
  2009.01858}}].

\bibitem{ref:smith_et_al_03}
R.~E. {Smith} et~al., \emph{{Stable clustering, the halo model and non-linear
  cosmological power spectra}},
  \href{https://doi.org/10.1046/j.1365-8711.2003.06503.x}{\emph{\mnras}
  {\bfseries 341} (2003) 1311}
  [\href{https://arxiv.org/abs/astro-ph/0207664}{{\ttfamily
  astro-ph/0207664}}].

\bibitem{ref:lewis_challinor_lasenby_99}
A.~{Lewis}, A.~{Challinor} and A.~{Lasenby}, \emph{{Efficient Computation of
  Cosmic Microwave Background Anisotropies in Closed Friedmann-Robertson-Walker
  Models}}, \href{https://doi.org/10.1086/309179}{\emph{\apj} {\bfseries 538}
  (2000) 473}.

\bibitem{ref:modi_et_al_17}
C.~{Modi}, M.~{White} and Z.~{Vlah}, \emph{{Modeling CMB lensing cross
  correlations with CLEFT}},
  \href{https://doi.org/10.1088/1475-7516/2017/08/009}{\emph{\jcap} {\bfseries
  2017} (2017) 009} [\href{https://arxiv.org/abs/1706.03173}{{\ttfamily
  1706.03173}}].

\bibitem{ref:healpy_paper}
A.~Zonca, L.~Singer, D.~Lenz, M.~Reinecke, C.~Rosset, E.~Hivon et~al.,
  \emph{healpy: equal area pixelization and spherical harmonics transforms for
  data on the sphere in python},
  \href{https://doi.org/10.21105/joss.01298}{\emph{Journal of Open Source
  Software} {\bfseries 4} (2019) 1298}.

\bibitem{ref:karamanis_beutler_21}
M.~{Karamanis} and F.~{Beutler}, \emph{{hankl: A lightweight Python
  implementation of the FFTLog algorithm for Cosmology}},
  \href{https://doi.org/10.48550/arXiv.2106.06331}{\emph{arXiv e-prints} (2021)
  arXiv:2106.06331} [\href{https://arxiv.org/abs/2106.06331}{{\ttfamily
  2106.06331}}].

\bibitem{ref:fang_et_al_20}
X.~{Fang}, E.~{Krause}, T.~{Eifler} and N.~{MacCrann}, \emph{{Beyond Limber:
  efficient computation of angular power spectra for galaxy clustering and weak
  lensing}}, \href{https://doi.org/10.1088/1475-7516/2020/05/010}{\emph{\jcap}
  {\bfseries 2020} (2020) 010}
  [\href{https://arxiv.org/abs/1911.11947}{{\ttfamily 1911.11947}}].

\bibitem{Lahav94}
O.~{Lahav}, K.~B. {Fisher}, Y.~{Hoffman}, C.~A. {Scharf} and S.~{Zaroubi},
  \emph{{Wiener Reconstruction of All-Sky Galaxy Surveys in Spherical
  Harmonics}}, \href{https://doi.org/10.1086/187244}{\emph{\apjl} {\bfseries
  423} (1994) L93} [\href{https://arxiv.org/abs/astro-ph/9311059}{{\ttfamily
  astro-ph/9311059}}].

\bibitem{Fisher94}
K.~B. {Fisher}, C.~A. {Scharf} and O.~{Lahav}, \emph{{A spherical harmonic
  approach to redshift distortion and a measurement of Omega(0) from the 1.2-Jy
  IRAS Redshift Survey}},
  \href{https://doi.org/10.1093/mnras/266.1.219}{\emph{\mnras} {\bfseries 266}
  (1994) 219} [\href{https://arxiv.org/abs/astro-ph/9309027}{{\ttfamily
  astro-ph/9309027}}].

\bibitem{HT95}
A.~F. {Heavens} and A.~N. {Taylor}, \emph{{A spherical harmonic analysis of
  redshift space}},
  \href{https://doi.org/10.1093/mnras/275.2.483}{\emph{\mnras} {\bfseries 275}
  (1995) 483} [\href{https://arxiv.org/abs/astro-ph/9409027}{{\ttfamily
  astro-ph/9409027}}].

\bibitem{Percival04}
W.~J. {Percival}, D.~{Burkey}, A.~{Heavens}, A.~{Taylor}, S.~{Cole}, J.~A.
  {Peacock} et~al., \emph{{The 2dF Galaxy Redshift Survey: spherical harmonics
  analysis of fluctuations in the final catalogue}},
  \href{https://doi.org/10.1111/j.1365-2966.2004.08146.x}{\emph{\mnras}
  {\bfseries 353} (2004) 1201}
  [\href{https://arxiv.org/abs/astro-ph/0406513}{{\ttfamily
  astro-ph/0406513}}].

\bibitem{Pad01}
N.~{Padmanabhan}, M.~{Tegmark} and A.~J.~S. {Hamilton}, \emph{{The Power
  Spectrum of the CFA/SSRS UZC Galaxy Redshift Survey}},
  \href{https://doi.org/10.1086/319700}{\emph{\apj} {\bfseries 550} (2001) 52}
  [\href{https://arxiv.org/abs/astro-ph/9911421}{{\ttfamily
  astro-ph/9911421}}].

\bibitem{Pratten13}
G.~{Pratten} and D.~{Munshi}, \emph{{Effects of linear redshift space
  distortions and perturbation theory on BAOs: a 3D spherical analysis}},
  \href{https://doi.org/10.1093/mnras/stt1854}{\emph{\mnras} {\bfseries 436}
  (2013) 3792} [\href{https://arxiv.org/abs/1301.3673}{{\ttfamily 1301.3673}}].

\bibitem{ref:samushia_19}
L.~{Samushia}, \emph{{Proper Fourier decomposition formalism for cosmological
  fields in spherical shells}},
  \href{https://doi.org/10.48550/arXiv.1906.05866}{\emph{arXiv e-prints} (2019)
  arXiv:1906.05866} [\href{https://arxiv.org/abs/1906.05866}{{\ttfamily
  1906.05866}}].

\bibitem{ref:gebhardt_dore_21}
H.~S. {Grasshorn Gebhardt} and O.~{Dor{\'e}}, \emph{{Fabulous code for
  spherical Fourier-Bessel decomposition}},
  \href{https://doi.org/10.1103/PhysRevD.104.123548}{\emph{\prd} {\bfseries
  104} (2021) 123548} [\href{https://arxiv.org/abs/2102.10079}{{\ttfamily
  2102.10079}}].

\bibitem{ref:passaglia_et_al_17}
S.~{Passaglia}, A.~{Manzotti} and S.~{Dodelson}, \emph{{Cross-correlating 2D
  and 3D galaxy surveys}},
  \href{https://doi.org/10.1103/PhysRevD.95.123508}{\emph{\prd} {\bfseries 95}
  (2017) 123508} [\href{https://arxiv.org/abs/1702.03004}{{\ttfamily
  1702.03004}}].

\bibitem{ref:hamilton_00}
A.~J.~S. {Hamilton}, \emph{{Uncorrelated modes of the non-linear power
  spectrum}},
  \href{https://doi.org/10.1046/j.1365-8711.2000.03071.x}{\emph{\mnras}
  {\bfseries 312} (2000) 257}
  [\href{https://arxiv.org/abs/astro-ph/9905191}{{\ttfamily
  astro-ph/9905191}}].

\bibitem{ref:loverde_afshordi_08}
M.~{LoVerde} and N.~{Afshordi}, \emph{{Extended Limber approximation}},
  \href{https://doi.org/10.1103/PhysRevD.78.123506}{\emph{\prd} {\bfseries 78}
  (2008) 123506} [\href{https://arxiv.org/abs/0809.5112}{{\ttfamily
  0809.5112}}].

\end{thebibliography}\endgroup
\end{document}